\documentclass[authoryear,12pt]{article}
\usepackage{amsmath}
\usepackage{times}
\usepackage{graphicx}
\usepackage{color}
\usepackage{multirow}
\usepackage{setspace}
\usepackage[authoryear]{natbib}
\usepackage{rotating}
\usepackage{bbm}
\usepackage{latexsym}
\usepackage[english]{babel}

\usepackage{subfig}
\usepackage{caption}
\usepackage{amssymb}
\usepackage{epsfig}

\begin{document}

\begin{titlepage}
\begin{center}
  {\Large The identity of information: how deterministic dependencies constrain information synergy and redundancy}
  \vspace*{3mm}

        {\large Daniel Chicharro $^{1,2,\ast}$, Giuseppe Pica$^{2}$, Stefano Panzeri$^{2}$}

        \vspace*{2mm}
        {$^1$ \emph{Department of Neurobiology, Harvard Medical School, Boston, MA, USA}}\\[1.5mm]

        {$^2$ \emph{Neural Computation Laboratory, Center for Neuroscience and Cognitive Systems@UniTn, Istituto Italiano di Tecnologia, Rovereto (TN), Italy.}}\\[1.5mm]

        {$\ast$ daniel\_chicharro@hms.harvard.edu or daniel.chicharro@iit.it}\\[1.5mm]

\end{center}

\vspace*{0mm}

\begin{center}
{\large Abstract}
\end{center}

Understanding how different information sources together transmit information is crucial in many domains. For example, understanding the neural code requires characterizing how different neurons contribute unique, redundant, or synergistic pieces of information about sensory or behavioral variables. Williams and Beer (2010) proposed a partial information decomposition (PID) which separates the mutual information that a set of sources contains about a set of targets into nonnegative terms interpretable as these pieces. Quantifying redundancy requires assigning an identity to different information pieces, to assess when information is common across sources. Harder et al.\,(2013) proposed an identity axiom stating that there cannot be redundancy between two independent sources about a copy of themselves. However, Bertschinger et al.\,(2012) showed that with a deterministically related sources-target copy this axiom is incompatible with ensuring PID nonnegativity. Here we study systematically the effect of deterministic target-sources dependencies. We introduce two synergy stochasticity axioms that generalize the identity axiom, and we derive general expressions separating stochastic and deterministic PID components. Our
analysis identifies how negative terms can originate from deterministic dependencies and shows how different assumptions on information identity, implicit in the stochasticity and identity axioms, determine the PID structure. The implications for studying neural coding are discussed.

\vspace*{3mm}


\end{titlepage}



\thispagestyle{empty}
%
\ \vspace{-0mm}\\

\pagenumbering{arabic}

\section{Introduction}
\label{s1}

The characterization of dependencies between the parts of a multivariate system helps to understand its function and its underlying mechanisms. Within the information-theoretic framework, this problem can be investigated by breaking down into parts the joint entropy of a set of variables \citep{Amari01, Schneidman03b, Ince10} or the mutual information between sets of variables \citep{Panzeri99b, Chicharro14b, Timme14}. These approaches have many applications to study dependencies in complex systems such as genes networks \citep[e.\,g.\,][]{Watkinson09, Erwin09, Chatterjee16}, neural coding and communication \citep[e.\,g.\,][]{Panzeri08, Marre09, Faes16}, or interactive agents \citep[e.\,g.\,][]{Katz11, Flack2012, Ay2012}.

An important aspect of how information is distributed across a set of variables concerns whether different variables provide redundant, unique or synergistic information when combined with other variables. Intuitively, variables share redundant information if each variable carries individually the same information carried by other variables. Information carried by a certain variable is unique if it is not carried by any other variables or their combination, and a group of variables carries synergistic information if some information arises only when they are combined. The presence of these different types of information has implications for example to determine how the information can be decoded \citep{Latham05}, how robust it is to disruptions of the system \citep{Rauh14b}, or how the variables’ set can be compressed without an information loss \citep{Tishby99}.

Characterizing the distribution of redundant, unique, and synergistic information is especially relevant in systems neuroscience, to understand how information is distributed in neural population responses. This requires identifying the features of neural responses that represent sensory stimuli and behavioural actions \citep{Averbeck06, Panzeri17} and how this information is transmitted and transformed across brain areas \citep{Wibral14, Timme16}. The breakdown of information into these different types of components can determine the contribution of different classes of neurons, and of different spatiotemporal components of population activity \citep{Panzeri10, Panzeri15}. Moreover, the identification of synergistic or redundant components of information transfer may help to map dynamic functional connectivity and integration of information across neurons or networks \citep{Valdes11, Vicente10, Ince15, Deco15}.

Despite the notions of redundant, unique, and synergistic information seem at first intuitive, their rigorous quantification within the information-theoretic framework has proven to be elusive. Synergy and redundancy have traditionally been quantified with the measure called interaction information \citep{McGill54} or co-information \citep{Bell03}, but this measure does not quantify them separately, and the presence of one or the other is associated with positive or negative values, respectively. Synergy has also been quantified using maximum entropy models as the information that can only be retrieved from the joint distribution of the variables \citep{Amari01, Olbrich15, Perrone16}.

However, a recent seminal work of \cite{Williams10} introduced a framework, called Partial Information Decomposition (PID), to more precisely and simultaneously quantify the redundant, unique, and synergistic information that a set of variables (or primary sources) S has about a target X. This decomposition has two cornerstones. The first is the definition of a general measure of redundancy following a set of axioms that impose desirable properties, in agreement with the corresponding abstract notion of redundancy \citep{Williams10b}. The second is the construction of a redundancy lattice, structured according to these axioms, which reflects a partial ordering of redundancies for different sets of variables \citep{Williams10}.

The PID framework has been adopted and further developed by many others \citep[e.\,g.\,][]{Harder12, Bertschinger12, Griffith13, Ince16, Rauh17, Chicharro17b}. However, its concrete implementation and the properties that the PID terms should have continue to be debated \citep{Rauh17b, Ince16}. \cite{Harder12} argued that the original redundancy measure of \cite{Williams10} quantifies only quantitatively equal amounts of information and not information that is qualitatively the same. They introduced a new axiom, namely the identity axiom, which states that when the target is a copy of two sources redundancy should correspond to the mutual information between them, and cancel for independent sources. Several measures that fulfill the identity axiom have been subsequently proposed in substitution of the original redundancy measure \citep{Harder12, Griffith13, Bertschinger12}. However, \cite{Bertschinger12b} provided a counterexample illustrating that in the multivariate case the identity axiom is incompatible with ensuring the nonnegativity of the PID terms. Like the target-source copy example used to motivate the axiom, also this counterexample involves deterministic target-sources dependencies.

Here we study in a general way the effect of deterministic target-sources dependencies in the PID decomposition. While the counterexample of \cite{Bertschinger12b} reveals the inconsistency of nonnegativity and the identity axiom, it does not provide a general clue of why they are incompatible and what has to be modified. Furthermore, while the identity axiom was advocated based on a concrete example, the question of generally determining the identity of different pieces of information has not been addressed independently of proposing specific redundancy measures. As we show in what follows, our analysis addresses more generally and explicitly the question of assigning information identity and the cause of negative PID terms.

We start this work reviewing the PID decompositions (Section \ref{s2}). We then introduce two alternative forms, a weak and a strong form, of a stochasticity axiom that imposes constraints to the existence of synergistic information in the presence of deterministic target-source dependencies (Section \ref{s3}). Using these axioms, we derive general expressions that separate each PID term into a stochastic and a deterministic component for the bivariate (Section \ref{s41}) and trivariate (Section \ref{s51}) case. We show how these axioms lead to two alternative generalizations of the identity axiom (Section \ref{s42}) and check if several previously proposed redundancy measures conform to these generalizations (Section \ref{s43}). We reconsider the examples used by \cite{Bertschinger12b}, characterizing their bivariate and trivariate decompositions and illustrating how in general negative PID terms can occur (Sections \ref{s44} and \ref{s52}). Finally, comparing the stochasticity and identity axioms, we discuss the implications of assuming a certain criterion to identify pieces of information in the target in the presence of deterministic target-sources dependencies, and concretely of assuming that their identity is related to specific sources (Section \ref{s45} and Section \ref{s53}).

\section{A review of the PID framework}
\label{s2}

The seminal work of \cite{Williams10} introduced a new approach to decompose the mutual information into a set of nonnegative contributions. Let us consider first the bivariate case. Assume that we have a target $X$ formed by one variable or by a set of variables and two variables $1$ and $2$ which information about $X$ we want to characterize. \cite{Williams10} argued that the mutual information of each variable can be expressed as

\begin{equation}
I(X;1) = I(X;1.2)+I(X;1 \backslash 2),
\label{e1}
\end{equation}
and similarly for $I(X;2)$. The term $I(X;1.2)$ refers to a redundancy component between variables $1$ and $2$, which can be obtained either by knowing $1$ or $2$ separately. The terms $I(X;1 \backslash 2)$ and $I(X;2 \backslash 1)$ quantify a component that is unique of $1$ and of $2$, respectively, that is, the information that can be obtained from one of the variables alone but that cannot be obtained from the other alone. Furthermore, the joint information of $12$ can be expressed as

\begin{equation}
I(X;12) =I(X;1.2)+I(X;1 \backslash 2)+I(X;2 \backslash 1)+I(X;12 \backslash 1,2),
\label{e2}
\end{equation}
where the term $I(X;12 \backslash 1,2)$ refers to the synergistic information of the two variables, which is unique for the joint source $12$ with respect to both variables alone. Therefore, given the standard information-theoretic chain rule equalities \citep{Cover06}

\begin{subequations}
\begin{align}
I(X;12) &= I(X;1)+I(X;2|1)\\
&= I(X;2)+I(X;1|2),
\end{align}
\label{e3}
\end{subequations}
the conditional mutual information is decomposed as

\begin{equation}
I(X;2|1) = I(X;2 \backslash 1)+I(X;12 \backslash 1,2),
\label{e4}
\end{equation}
and analogously for $I(X;1|2)$. Conditioning removes the redundant component but adds the synergistic component so that conditional information is the sum of the unique and synergistic terms.

In this decomposition a redundancy and a synergy component can exist simultaneously. In fact, \cite{Williams10} showed that the measure of co-information \citep{Bell03} that previously had been used to quantify synergy and redundancy, defined as

\begin{equation}
C(X;1;2) = I(i;j)-I(i;j|k) = I(i;j)+I(i;k)-I(i;j,k)
\label{e5}
\end{equation}
for any assignment of $\{X,1,2\}$ to $\{i,j,k\}$, corresponds to the difference between the redundancy and the synergy terms of Eq.\,\ref{e2}:

\begin{equation}
C(X;1;2) = I(X;1.2)-I(X;12 \backslash 1,2).
\label{e5_2}
\end{equation}

More generally, \cite{Williams10} defined decompositions of the mutual information about a target $X$ for any multivariate set of variables $S$. This general formulation relies on the definition of a general measure of redundancy and the construction of a redundancy lattice. In more detail, to decompose the information $I(X;S)$, \cite{Williams10} defined a \emph{source} $A$ as a subset of the variables in $S$, and a \emph{collection} $\alpha$ as a set of sources. They then introduced a measure of redundancy to quantify for each collection the redundancy between the sources composing the collection, and constructed a redundancy lattice which reflects the relation between the redundancies of all different collections. Here we will generically refer to the redundancy of a collection $\alpha$ by $I(X;\alpha)$. Furthermore, following \cite{Chicharro17}, we use a more concise notation than in \cite{Williams10}: For example, instead of writing $\{1\}\{23\}$ for the collection composed by the source containing variable $1$ and the source containing variables $2$ and $3$, we write $1.23$, that is, we save the curly brackets that indicate for each source the set of variables and we use instead a dot to separate the sources. We will also refer to the single variables in $S$ as \emph{primary} sources when we want to specifically distinguish them from general sources that can contain several variables.

\cite{Williams10b} argued that a measure of redundancy should comply with the following axioms:

\begin{itemize}
  \item \textbf{Symmetry}: $I(X;\alpha)$ is invariant to the order of the sources in the collection.
  \item \textbf{Self-redundancy}: The redundancy of a collection formed by a single source is equal to the mutual information of that source.
  \item \textbf{Monotonicity}: Adding sources to a collection can only decrease the redundancy of the resulting collection, and redundancy is kept constant when adding a superset of any of the existing sources.
\end{itemize}

\begin{figure}[!htbp]
  \begin{center}
    \scalebox{0.4}{\includegraphics*{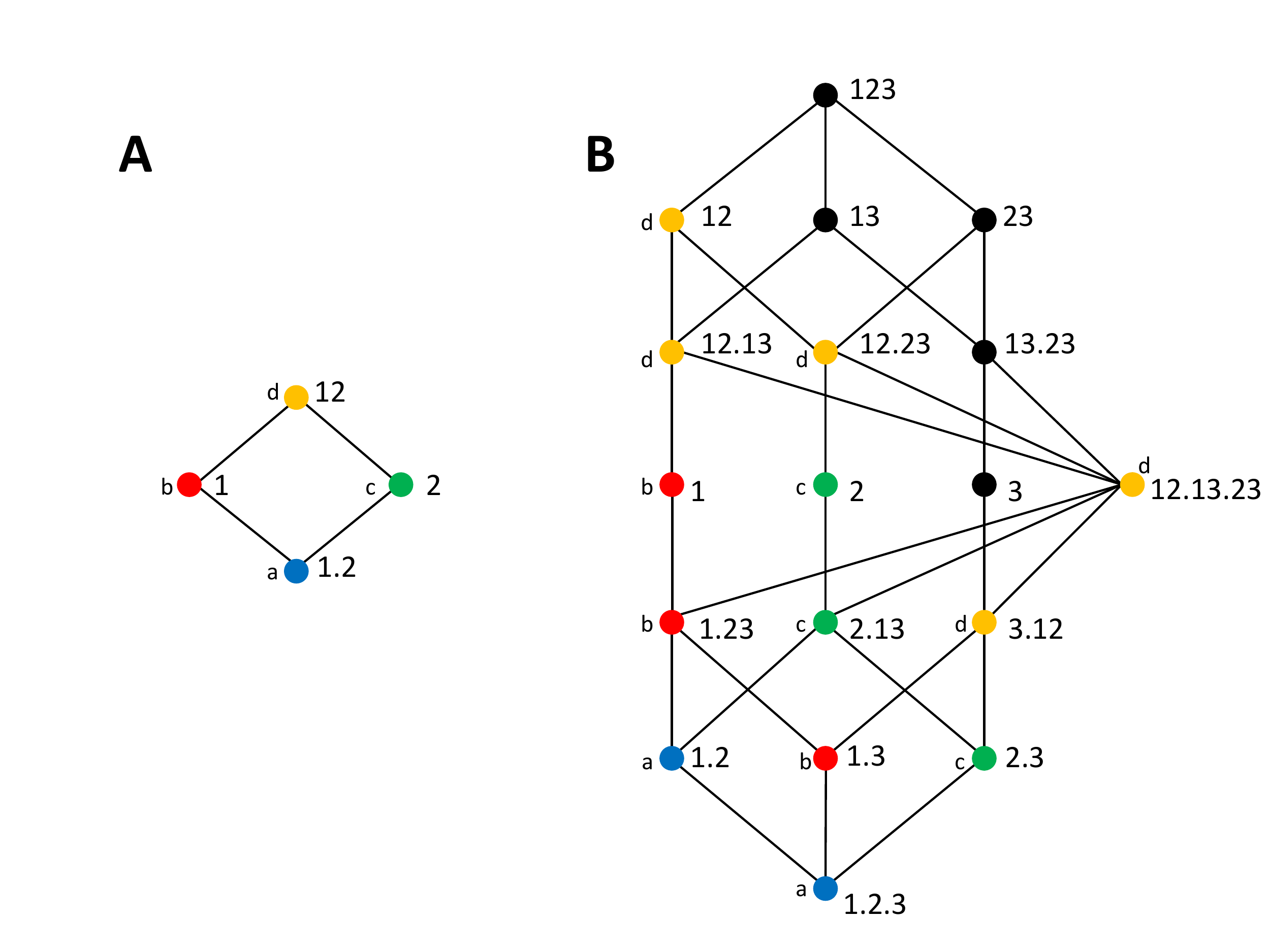}}
  \end{center}
  \caption{Redundancy lattices of \cite{Williams10}. The lattices reflect the partial ordering defined by Eq.\,\ref{e7}. \textbf{A}) Bivariate lattice corresponding to the decomposition of $I(X; 12)$. \textbf{B}) Trivariate lattice corresponding to the decomposition of $I(X; 123)$. The color and label of the nodes indicate the mapping of PID terms from the trivariate to the bivariate lattice, in particular nodes with the same color in the trivariate lattice are accumulated in the corresponding node in the bivariate lattice.}
  \label{fig1}
\end{figure}

The monotonicity property allows introducing a partial ordering between the collections, which is reflected in the redundancy lattice. Self-redundancy links the lattice to the joint mutual information $I(X;S)$ because at its top there is the collection formed by a single source including all the variables in $S$. Furthermore, the number of collections to be included in the lattice is limited by the fact that adding a superset of any source does not change redundancy. For example, the redundancy between the source $12$ and the source $2$ is all the information $I(X;2)$. Accordingly, the set of collections that can be included in the lattice is defined as

\begin{equation}
\mathcal{A}(S)= \{ \alpha \in \mathcal{P}(S) \backslash \{ \emptyset \}: \forall\ A_i, A_j \in \alpha, A_i \nsubseteq A_j\},
\label{e6}
\end{equation}
where $\mathcal{P}(S) \backslash \{ \emptyset \}$ is the set of all nonempty subsets of the set of nonempty sources that can be formed from $S$. This domain reflects the symmetry axiom in that it does not distinguish the order of the sources. For this set of collections, \cite{Williams10} defined a partial ordering relation to construct the lattice:

\begin{equation}
\forall\ \alpha, \beta \in \mathcal{A}(S), (\alpha \preceq \beta \Leftrightarrow \forall B \in \beta, \exists A \in \alpha, A \subseteq B),
\label{e7}
\end{equation}
that is, for two collections $\alpha$ and $\beta$, $\alpha \preceq \beta$ if for each source in $\beta$ there is a source in $\alpha$ that is a subset of that source. This partial ordering relation is reflexive, transitive, and antisymmetric. In fact, the consistency of the redundancy measures with the partial ordering of the collections, that is, that $I(X;\alpha) \leq I(X;\beta)$ iif $\alpha \preceq \beta$, represents a stronger form of the monotonicity axiom.

The mutual information multivariate decomposition was constructed in \cite{Williams10} by implicitly defining partial information measures $\Delta(X;\alpha)$ associated with each node $\alpha$ of the redundancy lattice, such that redundancy measures are obtained from the sum of partial information measures:

\begin{equation}
I(X;\alpha) = \sum_{\beta \in \downarrow \alpha} \Delta(X;\beta),
\label{e8}
\end{equation}
where $\downarrow\alpha$ refers to the set of collections lower than or equal to $\alpha$ in the partial ordering, and hence reachable descending from $\alpha$ in the lattice. The partial information measures are obtained inverting Eq.\,\ref{e8} by applying the principle of inclusion-exclusion to the terms in the lattice \citep{Williams10}. Redundancy lattices for $S$ being bivariate and trivariate are shown in Figure \ref{fig1}. As studied in \cite{Chicharro17}, a mapping exists between the terms of the trivariate and bivariate PID decompositions, as indicated by the colors and labels.

An extra axiom, called the identity axiom, was later introduced by \cite{Harder12} specifically for the bivariate redundancy measure:

\begin{itemize}
  \item \textbf{Identity axiom:} For two sources $A_1$ and $A_2$,  $I(A_1 \cup A_2; A_1.A_2)$ is equal to $I(A_1;A_2)$.
\end{itemize}

\cite{Harder12} pointed out that with the original measure of redundancy of \cite{Williams10} a nonzero redundancy is obtained for two independent variables and a target being a copy of them, and that a measure quantifying the amount of qualitatively common information and not the quantitatively equal amount of information should be zero in this case. \cite{Ince16} has specifically differentiated between the identity axiom, which assumes the form of redundancy for any degree of dependence between the primary sources when the target is a copy of them, and a more concrete property, namely the Independent Identity property, which only requires that redundancy about the copy target cancels when the primary sources are independent. Several alternative measures have been proposed that fulfill this additional axiom \citep{Harder12, Bertschinger12, Griffith13}. The properties of the PID terms have been characterized, either based on the axioms and the structure of the redundancy lattice \citep{Chicharro17, Pica17}, or also considering the properties of specific measures \citep{Bertschinger12b, Griffith13, Rauh14, Griffith14, Banerjee15, Rauh17b}. However, only for specific cases such as multivariate Gaussian systems with univariate targets, it has been shown that several of the proposed measures are actually equivalent \citep{Barret15, Faes17}.

\section{Stochasticity axioms for synergistic information}
\label{s3}

We start our analysis of deterministic relations between the target $X$ and the set of primary sources $S$ by enunciating two versions of a stochasticity axiom for synergistic information. These axioms impose different constraints on the synergy terms when the dependency of the target upon the sources can be partly deterministic and partly stochastic. We consider first the weak axiom. This axiom is motivated by the idea that if any subset $X'$ of variables comprised in the target $X$ can be completely determined by a source corresponding to a subset $S'$ of $S$ then there cannot be synergistic information about that subset $X'$ between $S'$ and any other sources. This is because $S'$ can already provide all the information about $X'$ without combining it with any other variable. Accordingly, the weak axiom assumes that:

\vspace*{3mm}

\noindent \textbf{Synergy weak stochasticity axiom:} \emph{For a target $X$ and a set of variables $S$, if there is a subset $X'= X(S')$ of $X$ such that it can be determined completely from a subset $S'$ of $S$, then}

\begin{equation}
\Delta(X; \alpha) = \Delta(X \backslash X(S'); \alpha)\ \ \forall \alpha \notin \bigcup_{i \in S} \downarrow i,
\label{r0}
\end{equation}
\emph{where $\downarrow i$ indicates the collections reachable by descending the lattice from node $i$, corresponding to a primary source.}

\vspace*{2mm}

That is, any synergistic term about $X$ is equal to the synergy about a target $X \backslash X(S')$ that does not include the variables $X(S')$ determined by $S'$.

This weak form of the stochasticity axiom implies that the sources cannot have synergistic information about a part $X’$ of the target that is deterministically related to them. However, the axiom does not restrict that those variables in $S'$ that determine $X'$ may provide information about other parts of the target in a synergistic way. Conversely, a strong form of the stochasticity axiom imposes that the variables in $S'$ can only provide synergistic information to the degree that they are not themselves deterministically related to the variables in $X'$. In particular, it assumes that:

\vspace*{3mm}

\noindent \textbf{Synergy strong stochasticity axiom:} \emph{For a target $X$ and a set of variables $S$, if there is a subset $X' = X(S')$ of $X$ such that it can be determined completely from a subset $S'$ of $S$, then}

\begin{equation}
\Delta(X; \alpha) = \Delta(X \backslash X(S'); \alpha | X(S'))\ \ \forall \alpha \notin \bigcup_{i \in S} \downarrow i.
\label{r0b}
\end{equation}

\vspace*{2mm}

That is, the synergy about $X$ is equal to the synergy in the lattice associated with the decomposition of the mutual information $I(X \backslash X(S'); S |X(S'))$ that $S$ has about $X \backslash X(S')$ conditioned on $X(S')$. The logic of the strong axiom can be better appreciated when, instead of just a functional relation, some of the primary sources are themselves contained in the target (i.\,e.\,$X(S') = S'$). In this case the strong axiom states that there cannot be any synergistic contribution involving variables in $S'$. In contrast to the weak axiom, these contributions cannot be present even if providing information about $X \backslash X(S')$. The motivation is that the primary sources in $S'$ cannot provide other information about the target than the information about themselves, which can be provided without combining them with any other variable. Accordingly, when $X'=S'$,

\begin{equation}
\Delta(X; \alpha) = 0 \ \ \forall \alpha \notin \bigcup_{i \in S} \downarrow i\ : \exists A \in \alpha, S' \cap A \neq \emptyset,
\label{r1b}
\end{equation}
that is, there is no synergy for those nodes whose collection has a source containing a variable from $S'$.

In this work we will study how, based on these axioms, bivariate and trivariate PID decompositions are affected by deterministic relations between the target and the primary sources. To simplify the derivations we will focus on the case in which the target $X$ contains some of the primary sources themselves. A more general formulation that considers target variables determined as a function of the sources leads to the same main qualitative conclusions. All the derivations follow from the relations characteristic of the redundancy lattice, and we do not need to select any specific measure of redundant, unique or synergistic information.

\section{Bivariate decompositions with deterministic target-sources dependencies}
\label{s4}

We start with the bivariate case. Consider that the target $X$ may have some overlap $X \cap 12$ with the sources $1$ and $2$. Following the weak stochasticity axiom (Eq.\,\ref{r0}) synergy is expressed as:

\begin{equation}
I(X; 12 \backslash 1,2) = I(X \backslash 12 ; 12 \backslash 1,2).
\label{r2}
\end{equation}
On the other hand, for the strong stochasticity axiom (Eq.\,\ref{r1b}) we have:
\begin{equation}
I(X; 12 \backslash 1,2) = \begin{cases} I(X \backslash 12; 12 \backslash 1,2) \ \ \mathrm{if}\ X \cap 12 = \emptyset \\ 0 \ \ \mathrm{if}\ X \cap 12 \neq \emptyset\end{cases}.
\label{r2b}
\end{equation}

Given these expressions of the synergistic terms we will now derive how deterministic relations affect the other PID terms.

\subsection{General formulation}
\label{s41}

For both forms of the stochasticity axiom we will derive expressions of unique and redundant information in the presence of a target-sources overlap. These derivations follow the same procedure: First, given that unique and synergistic information are related to conditional mutual information by Eq.\,\ref{e4}, the synergy stochasticity axioms determine the form of the unique information terms. Second, once the unique information terms are derived, their relation to the mutual information together with the redundancy term (Eq.\,\ref{e1}) allows identifying redundancy. For both unique and redundant information terms this procedure separates stochastic and deterministic components. However, how these components are combined depends on the order in which stochastic and deterministic target-sources dependencies are partitioned. In particular, using the chain rule \citep{Cover06} of the mutual information we can separate the information about the target in two different ways:

\begin{subequations}
\begin{align}
I(X;12) &= I(X \backslash 12; 12) + I(X \cap 12 ; 12| X \backslash 12) \\
&= I(X \cap 12; 12) + I(X \backslash 12 ; 12| X \cap 12).
\end{align}
\label{r3}
\end{subequations}
The first case considers first the stochastic dependencies and after the conditional deterministic dependencies. In the second case, this order is reversed. We will see that for each axiom only one of these partitioning orders leads to expressions that additively separate stochastic and deterministic components for each PID terms.

\subsubsection{PID decompositions with the weak axiom}
\label{s411}

We start with the PID decomposition of $I(X;12)$ derived from the weak axiom (Eq.\,\ref{r2}). Consider the mutual information partitioning order of Eq.\,\ref{r3}a, which can be reexpressed as

\begin{equation}
\begin{split}
I(X;12) = I(X \backslash 12; 12) + H(X \cap 12| X \backslash 12),
\label{r6}
\end{split}
\end{equation}
that is, the second summand corresponds to the conditional entropy of the overlapping target variables given the non-overlapping ones. We now proceed analogously for the PID terms. Since conditional mutual informations are the sum of a unique and a synergistic information component (Eq.\,\ref{e4}), we have that

\begin{equation}
\begin{split}
&I(X; 1 \backslash 2) = I(X; 1|2) - I(X; 12 \backslash 1,2)\\
&= I(X \backslash 12 ;1|2) + I(X \cap 12; 1|2, X \backslash 12)- I(X \backslash 12; 12 \backslash 1,2).
\label{r4}
\end{split}
\end{equation}
The first equality indicates that unique information is conditional information minus synergy. The second equality uses the chain rule to separate the conditional mutual information stochastic and deterministic components, and applies the stochasticity axiom to remove the overlapping part of the target in the synergy term. Using again the relation between conditional mutual information and unique and synergistic terms but now for the target $X \backslash 12$ we get

\begin{equation}
I(X; 1 \backslash 2) = I(X \backslash 12 ;1 \backslash 2)+ H(X \cap 1| 2, X \backslash 12),
\label{r5}
\end{equation}
where we also used that $I(X \cap 12; 1|2, X \backslash 12)$ equals the entropy $H(X \cap 1| 2, X \backslash 12)$. Accordingly, the unique information of $1$ can be separated into a stochastic component, the unique information about target $X \backslash 12$, and a deterministic component, the entropy $H(X \cap 1| 2, X \backslash 12)$. This last term is zero if the target does not contain source $1$. If it does, it quantifies the entropy that only $1$ as a source can explain about itself as part of the target, which is thus an extra unique information contribution.

Once we have identified the unique information stochastic and deterministic components we can use the relation of unique and redundant information with the mutual information (Eq.\,\ref{e1}) to characterize the redundancy. We get that:

\begin{equation}
I(X; 1.2) = I(X \backslash 12; 1.2) + \begin{cases} 0 \ \  \mathrm{if}\ X \cap 12 = \emptyset \\ I(1;2|X \backslash 12)\ \  \mathrm{if}\ X \cap 12 \neq \emptyset \end{cases}.
\label{r7}
\end{equation}
Therefore, it suffices that one of the two primary sources overlaps with the target so that their conditional mutual information given the non-overlapping target variables contributes to redundancy.

We can follow the same procedure to derive expressions for the unique and redundant information terms but
applying the other mutual information partitioning order of Eq.\,\ref{r3}b. The resulting terms can be compared in Table \ref{tab1} and are derived in more detail in Appendix \ref{a0}, where we also show the consistency between the expressions obtained with each partitioning order. In the upper part of the table we collect the decompositions into stochastic and deterministic contributions for each PID term and for the two partitioning orders. To simplify the expressions, their form is shown only for the case of $X \cap i \neq \emptyset$. With the alternative partitioning order, both the expressions of unique information and redundancy contain a cross-over component, namely the synergy about $X \backslash 12$, instead of being expressed in terms of the unique information and redundancy of $X \backslash 12$, respectively. Furthermore, the separation of the deterministic and stochastic components is not additive. This indicates that, while the chain rule holds for the mutual information, it is not guaranteed that the same type of separation holds separately for each PID term. Only for a certain partitioning order, when stochastic dependencies are considered first, unique and redundant information terms derived from the weak axiom can both be separated additively into a stochastic and a deterministic component without cross-over terms. In the lower part of the table we individuate the deterministic PID components obtained from the partitioning order for which each PID term is separated additively into a stochastic and deterministic component.

\begin{table}[!htbp]
\centering
\begin{tabular}{| l c | c |}
\hline \hline
Term &  Decomposition\\
\hline \hline
$I(X; ij \backslash i,j)$ & $I(X \backslash ij; ij \backslash i,j)$\\
\hline
$I(X; i \backslash j)$ & $\begin{array}{c} I(X \backslash ij; i \backslash j) + H(i|j, X \backslash ij)\\ H(i|j) - I(X \backslash ij; ij \backslash i,j)\end{array}$ \\
\hline
$I(X; i.j)$ & $\begin{array}{c} I(X \backslash ij; i.j) + I(i;j |X \backslash ij )\\ I(i;j) + I(X \backslash ij; ij \backslash i,j) \end{array} $ \\
\hline \hline
Term &  Measure\\
\hline \hline
$\Delta_d(X; ij)$ & $0$ \\
\hline
$\Delta_d(X; i)$ & $H(i|j, X \backslash ij)$ \\
\hline
$\Delta_d(X; i.j)$ & $I(i;j|X \backslash ij)$ \\
\hline \hline
\end{tabular}
\caption{Decompositions of synergistic, unique, and redundant information terms into stochastic and deterministic contributions obtained assuming the weak stochasticity axiom. For each term we show the decompositions resulting from two alternative mutual information partitioning orders (Eq.\,\ref{r3}), which are consistent with each other (see Appendix \ref{a0}). For the partitioning order leading to an additive separation of each PID term into a stochastic and deterministic component we also individuate the deterministic contributions $\Delta_d(X; \beta)$. Synergy has only a stochastic component, according to the axiom (Eq.\,\ref{r2}). Expressions of unique information come from Eqs.\,\ref{r5} and \ref{r9}, and the ones of redundancy from Eqs.\,\ref{r7} and \ref{r12}. The expressions have been simplified with respect to the equations, indicating their form for the case $X \cap i \neq \emptyset$. The terms $\Delta_d(X; \beta)$ have analogous expressions for $X \cap j \neq \emptyset$ when a symmetry exists between $i$ and $j$ and are zero otherwise.}
\label{tab1}
\end{table}

\subsubsection{PID decompositions with the strong axiom}
\label{s412}

The procedure to derive the unique and redundant PID terms is the same if the strong stochasticity axiom is assumed, but determining synergy with Eq.\,\ref{r2b} instead of Eq.\,\ref{r2}. To simplify the expressions we indicate in advance that if $X \cap 12 = \emptyset$ each PID term with target $X$ is by definition equal to the one with target $X \backslash 12$ and we only provide expressions derived with some target-sources overlap. In contrast to the weak axiom, with the strong axiom an additive separation of stochastic and deterministic components is obtained with the partitioning order of Eq.\,\ref{r3}b. See Appendix \ref{a0} for details about the other partitioning order. For the unique information we obtain
\begin{equation}
I(X; 1 \backslash 2) = \begin{cases} I(X \backslash 12 ;1 \backslash 2) + I(X \backslash 12; 12 \backslash 1, 2)\ \ \mathrm{if}\ X \cap 1 = \emptyset \\ H(1|2)\ \ \mathrm{if}\ X \cap 1 \neq \emptyset \end{cases},
\label{r31}
\end{equation}
and for the redundancy
\begin{equation}
I(X; 1.2) = I(1;2).
\label{r32}
\end{equation}
As before, we summarize the PID decompositions in Table \ref{tab1b}. Comparing Table \ref{tab1} and \ref{tab1b} we see that the expressions obtained with the weak and strong axiom differ because of a cross-over contribution, corresponding to the synergy about $X \backslash 12$, which is transferred from redundancy to unique information. This is due to the synergy constraints imposed by each axiom: the strong axiom assumes that there is no synergy, and hence this part of the information has to be transferred to the unique information because the sum of synergy and unique information is constrained to equal the conditional mutual information. As a consequence, redundancy is reduced by an equivalent amount to comply with the constraints that relate unique informations and redundancy to mutual informations. Furthermore, like for the weak axiom, the chain rule property does not generally hold for each PID term separately. PID terms are consistent with the mutual information decompositions obtained applying the chain rule, but depending on the partitioning order and on the version of the axiom assumed, information contributions are redistributed between different PID terms, and between their stochastic and deterministic components. This is in agreement with previous concrete counterexamples provided by \cite{Bertschinger12b} and \cite{Rauh14} that showed that the chain rule does not hold in general for each PID term.

\begin{table}[!htbp]
\centering
\begin{tabular}{| l c | c |}
\hline \hline
Term &  Decomposition\\
\hline \hline
$I(X; ij \backslash i,j)$ & $0$\\
\hline
$I(X; i \backslash j)$ & $\begin{array}{c} I(X \backslash ij; i \backslash j)+ I(X \backslash ij; ij \backslash i,j) + H(i|j, X \backslash ij)\\ H(i|j) \end{array}$ \\
\hline
$I(X; i.j)$ & $\begin{array}{c} I(i;j |X \backslash ij )+ I(X \backslash ij; i . j)- I(X \backslash ij; ij \backslash i, j)\\ I(i;j) \end{array} $ \\
\hline \hline
Term &  Measure\\
\hline \hline
$\Delta_d(X; ij)$ & $0$ \\
\hline
$\Delta_d(X; i)$ & $H(i|j)$ \\
\hline
$\Delta_d(X; i.j)$ & $I(i;j)$ \\
\hline \hline
\end{tabular}
\caption{Decompositions of synergistic, unique, and redundant information terms into stochastic and deterministic contributions obtained assuming the strong stochasticity axiom. The table is analogous to Table \ref{tab1}. Synergy cancels according to the axiom (Eq.\,\ref{r2b}). Expressions of unique information come from Eqs.\,\ref{r29} and \ref{r31}, and the ones of redundancy from Eqs.\,\ref{r30} and \ref{r32}. Again, expressions are shown for the case $X \cap i \neq \emptyset$, with the corresponding symmetries holding for $X \cap j \neq \emptyset$ and with terms $\Delta_d(X; \beta)$ equal to zero otherwise.}
\label{tab1b}
\end{table}

\subsection{The relation between the synergy stochasticity axioms and the redundancy identity axiom}
\label{s42}

The two forms of the stochasticity axiom result in different expressions for the redundancy term. We now examine how these expressions are related to the redundancy identity axiom \citep{Harder12}. This axiom determines redundancy for a very specific deterministic target-sources relation, namely when there are two primary sources $1$ and $2$ and the target is equal to them, $X = 12$. It is straightforward to see that the redundancy identity axiom is subsumed by both stochasticity axioms:

\vspace*{3mm}

\noindent \textbf{Proposition:} \emph{The fulfillment of the synergy weak or strong stochasticity axioms implies the fulfillment of the redundancy identity axiom}

\vspace*{2mm}

\emph{Proof}: If $X = 12$ then $X \cap 12 = 12$ and $X \backslash 12 = \emptyset$. For the weak stochasticity axiom, redundancy (Eq.\,\ref{r7}) reduces to $I(12; 1.2) = I(1;2)$. For the strong stochasticity axiom, Eq.\,\ref{r32} is already $I(12; 1.2) = I(1;2)$.                $\Box$

\vspace*{3mm}

Therefore, the stochasticity axioms represent two alternative generalizations of the redundancy identity axiom: First, they do not only consider a target that is a copy of the primary sources, but a target with any degree of overlap or functional dependence with the sources. Second, they are not restricted to the bivariate case but are formulated for any number of primary sources.

Redundancy terms derived from each axiom coincide for the particular case that is addressed by the identity axiom, but more generally differ. The strong axiom leads to redundancy being equal to $I(1;2)$ not only for the case addressed by the identity axiom but in general when $X \cap 12 \neq \emptyset$ and independently of which are the non-overlapping target variables. Conversely, with the weak axiom redundancy depends on these other variables. We will further discuss these differences below based on concrete examples.

\subsection{The compliance of the stochasticity axioms by concrete measures}
\label{s43}

We now check for several proposed measures if they conform to the predictions of the stochasticity axioms. In particular we examine the original redundancy measures of \cite{Williams10}, the one based on the pointwise common change in surprisal of \cite{Ince16}, and the one based on maximum entropy of \cite{Bertschinger12}.

It is well-known that the redundancy measure of \cite{Williams10} does not comply with the identity axiom \citep{Harder12}. Even if $I(1;2)=0$, a redundancy $I(12; 1.2)>0$ can be obtained. This excess of redundancy leads to less unique information, which in turn produces a nonzero synergistic contribution inconsistently with both the weak and strong stochasticity axioms. Neither the redundancy measure of \cite{Ince16} complies with the identity axiom, and thus it does not conform to the stochasticity axioms.

Conversely, the redundancy measure of \cite{Bertschinger12} fulfills the identity axiom. To see how more broadly it compares to the redundancies derived from the weak and strong axioms consider the following example: if there is a target $X = 23$ and two sources $1$ and $2$, according to the weak axiom (Table \ref{tab1}) the redundancy $I(23; 1.2)$ should be equal to the sum of a deterministic component $I(1;2)$ and of a stochastic component $I(3; 12 \backslash 1,2)$. Conversely, according to the strong axiom, the redundancy equals only $I(1;2)$ (Eq.\,\ref{r32}). The redundancy measure of \cite{Bertschinger12} is calculated by minimizing the mutual information that the sources have about the target within the family of distributions which preserves the marginals of the target with each of the sources. In particular, redundancy is calculated as the co-information for the distributions leading to the minimal information within the family. In this example preserving p(1, 23), the marginal of the target $23$ and source $1$, implies preserving the whole joint distribution $p(1,2,3)$ and hence the minimal information within the family is equal to the original information. Accordingly, given Eq.\,\ref{e5}, for the \cite{Bertschinger12} measure

\begin{equation}
\begin{split}
I(23; 1.2) &= I(23;1)+I(23;2)-I(23;1,2)\\
&= I(23;1)+ H(2)-[H(2)+I(3;1|2)]\\
&= I(1;2).
\label{r14b}
\end{split}
\end{equation}
This redundancy measure coincides with the one predicted from the strong axiom. This holds in general, because it is a property of the co-information that if $X$ overlaps with $1$ or $2$ then $C(X;1;2) = I(1;2)$. Given this matching of the redundancy, it is straightforward to check that the rest of PID terms match as well.

\subsection{Illustrative systems}
\label{s44}

So far we have derived the predictions for the PID decompositions according to each version of the stochasticity axiom, pointed out the relation with the identity axiom, and checked how different previously proposed measures conform to these predictions. We now use concrete examples to further examine the decompositions. In particular, we reconsider two examples that have been previously studied in \cite{Bertschinger12b} and \cite{Rauh14}, namely the decompositions of the mutual information about a target jointly formed by the inputs and the output of a logical XOR operation or of an AND operation. We first describe below the decompositions obtained and in Section \ref{s45} we will discuss them in relation to underlying assumptions on how to assign an identity to different pieces of information of the target. The deterministic components for these examples are derived without assuming any specific measure of redundancy, unique, or synergistic information. The stochastic components have already been previously studied and some of the terms depend on the measures selected. We will indicate previous work examining these terms when required.

\begin{figure}[!htbp]
  \begin{center}
    \scalebox{0.4}{\includegraphics*{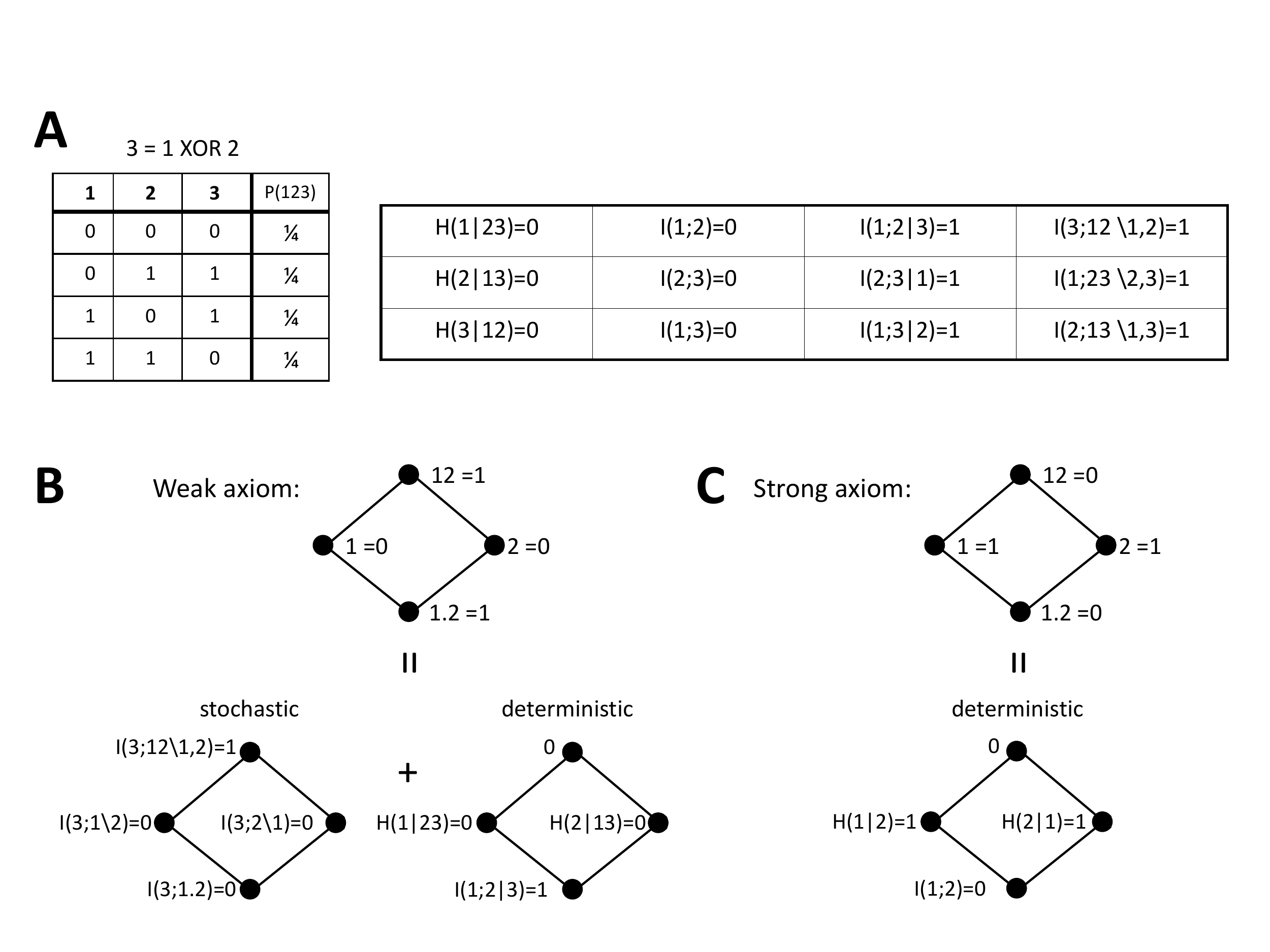}}
  \end{center}
  \caption{Bivariate decomposition of $I(123;12)$ for the XOR system. \textbf{A}) Joint distribution of the inputs $1$ and $2$ and the output $3$ for the XOR operation. We also collect the value of the information-theoretic quantities used to calculate this bivariate decomposition and the trivariate decomposition $I(123;123)$ in Section \ref{s52}. \textbf{B}) Bivariate decomposition derived from the weak stochasticity axiom. Stochastic and deterministic components are separated in agreement with Table \ref{tab1}. \textbf{C}) Bivariate decomposition derived from the strong axiom. Only deterministic components are present, following Table \ref{tab1b}.}
  \label{fig2a}
\end{figure}

\subsubsection{XOR}
\label{s441}

We start with the XOR operation. Consider an output variable $3$ determined through the operation $3=1\ \mathrm{XOR}\ 2$, resulting in the joint probability displayed in Figure \ref{fig2a}A. We also indicate the values of the information-theoretic measures needed to calculate the PID bivariate decompositions studied here and that will also serve for the trivariate decompositions addressed in Section \ref{s52}. We want to examine the decomposition of $I(123; 1,2)$, where the target is composed by the three variables. For each version of the stochasticity axiom we will focus on the mutual information partitioning order that allows separating additively a stochastic and a deterministic component of each PID term.

Since $X \backslash 12 =3$, for the weak axiom the PID decomposition (Figure \ref{fig2a}B) can be obtained by implementing the decomposition of $I(3; 12)$ and separately calculating the deterministic PID components $\Delta_d(123; \beta)$ as collected in Table \ref{tab1}. The decomposition of $I(3;12)$ for the XOR operation has been characterized repeatedly \citep[e.\,g.\,][]{Griffith13}, showing that all terms are zero except the synergy, which contributes one bit of information. There is no stochastic redundancy or unique information because $I(3;i)=0$ for $i=1,2$. Regarding the deterministic components, redundancy has $1$ bit because $I(1;2|3)=1$. The deterministic unique information components are zero because $H(i|jk)=0$ for $i=1,2$ and, according to the axiom, there is no deterministic synergy.

In the case of the strong axiom (Figure \ref{fig2a}C), since both primary sources overlap with the target, only deterministic components appear in the decomposition when selecting the partitioning order that additively separates stochastic and deterministic contributions, as indicated in Table \ref{tab1b}. By assumption, there is no synergy. Since $I(1;2)=0$, the redundancy is also zero and all the information is contained in the unique information terms. As pointed out for the generic expressions, the two decompositions differ in the transfer of the stochastic component of synergy to unique information, which in turns forces an equivalent transfer from redundancy to unique information.

We can compare these decompositions with previous analyses of this example \citep{Bertschinger12b, Rauh14}. In these studies the PID terms were derived using the identity axiom. In particular, it was argued that, since $12$ totally determines $3$, the target can be reduced from $123$ to $12$ and redundancy is thus $I(123; 1.2)= I(12;1.2)$. Then, using the identity axiom, $I(12;1.2)= I(1;2)$, which is zero. This reasoning leads to the same decomposition derived from the strong stochasticity axiom.

\subsubsection{AND}
\label{s442}

\begin{figure}[!htbp]
  \begin{center}
    \scalebox{0.4}{\includegraphics*{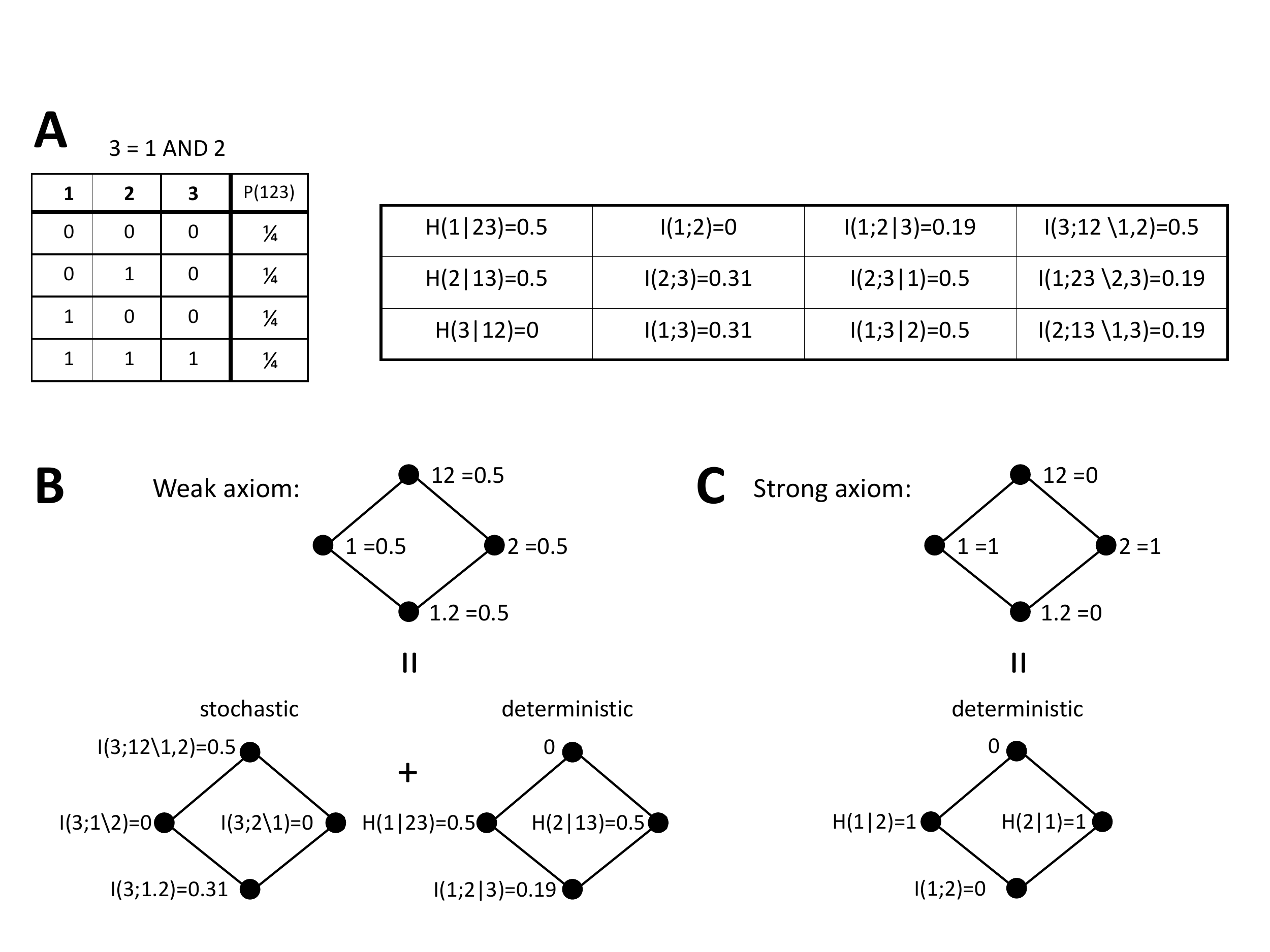}}
  \end{center}
  \caption{Bivariate decomposition of $I(123;12)$ for the AND system. The structure of the figure is analogous to Figure \ref{fig2a}.}
  \label{fig2b}
\end{figure}

As a second example, we now consider the AND operation. Following the weak axiom, again the decomposition can be obtained by implementing the PID decomposition of $I(3; 12)$ and separately calculating the deterministic PID components from Table \ref{tab1}, using the joint distribution of inputs and output displayed in Figure \ref{fig2b}A. The PID decomposition of $I(3; 12)$ for the AND operation has also been already characterized \citep[e.\,g.\,][]{Harder12}. For $I(123;12)$, each PID term contributes half a bit. Unique contributions come exclusively from the deterministic components. Each unique information has half a bit because the output and one input determine the other input only when not both have a value of $0$. Redundancy is also half a bit, but it comes in part from a stochastic component and in part from a deterministic one. The stochastic component was previously determined by \cite{Harder12}, indicating that this redundancy about the output appears intrinsically because of the AND mechanism, even if the inputs are independent. The deterministic component appears because, although the inputs are independent, conditioned on the output $I(1;2|3)>0$. The synergy $I(3; 12 \backslash 1, 2)=0.5$ was also previously determined by \cite{Harder12}. This PID decomposition differs from the one obtained with the weak axiom for the XOR example. Conversely, with the strong axiom the decomposition is the same as for the XOR example, because it is completely determined by $I(1;2)=0$. This latter decomposition is again in agreement with the arguments of \cite{Bertschinger12b} and \cite{Rauh14} based on the identity axiom.

\subsection{Implications of the stochasticity axioms for the notions of redundant, unique, and synergistic information}
\label{s45}

Each version of the stochasticity axiom implies a different quantification of redundancy. Since the value of unconditional and conditional mutual informations does not depend on the PID decomposition, for the bivariate case the extra constraints on synergy of the strong axiom imply assigning more information to unique information, which in turns restricts the amount of redundancy, as compared to the
constraints implied by the weak axiom. This restriction imposes that, if there is some target-sources overlap, redundancy only depends on the mutual information between the primary sources and is independent of dependencies between the sources and the other target variables.

We now examine in more detail how these different quantifications are related to the notion of redundancy as common information about the target that can be obtained by observing either source alone. The key point is how identity is assigned to different pieces of information in order to assess which information about the target carried by the sources is qualitatively common. In particular, for the strong axiom, its logic is that if a source is part of the target it cannot provide other information about the target than the information about itself and thus, if the other source does not contain information about it, this information is unique. Implicit in this argument there is the assumption that when a primary source is part of the target we can still identify and separate the bits of information about that source from the information about the rest of the target. This idea regarding the identity of the bits that are shared also motivated the introduction of the identity axiom. Although the identity axiom was formulated for sources with any degree of dependence, its motivation was mainly based on the case in which $I(1;2)=0$, when the sources are independent \citep{Harder12}. In this case, we can identify the bits of information related to variable $1$ and the ones to variable $2$, and thus redundancy, if it quantifies the qualitatively equal information that is shared and not only common amounts of information, has to cancel.

But assigning an identity to pieces of information in the target is in general less straightforward. For the XOR example, we will now consider different combinations of mutual information partitioning orders for $I(123;1)$ and $I(123;2)$ and show how, if the assignment of identity of the bits in the target $123$ is based on their association with the sources $1$ and $2$, the interpretation of redundant and unique information is ambiguous. First, consider that we decompose the information of each primary source as follows:

\begin{equation}
\begin{split}
I(123;1) &= I(1;1)+I(2;1|1)+I(3;1|12) = I(1;1) = H(1)\\ I(123;2) &= I(2;2)+ I(1;2|2)+I(3;2|12) = I(2;2)= H(2).
\label{r33}
\end{split}
\end{equation}
If we assume that we can identify the bit of information carried by each primary source about the target, these decompositions would suggest that there is no redundant information, since each source only carries one bit of information about itself and $I(1;2)=0$ for the XOR system. However, keeping the same decomposition of $I(123;1)$, we can consider alternative decompositions of $I(123;2)$:

\begin{subequations}
\begin{align}
I(123;2) &= I(3;2)+ I(1;2|3)+I(2;2|13) = I(1;2|3)= H(1) \\ &= I(1;2)+ I(3;2|1)+ I(2;2|13) = I(3;2|1) = H(3).
\end{align}
\label{r34}
\end{subequations}
The redundancy and unique information terms should not depend on how we apply the chain rule to $I(123;2)$. However, in contrast to Eq.\,\ref{r33}, the first decomposition of Eq.\,\ref{r34}a suggests, if the identity of the bits in the target is related to the overlapping variables in the sources, that there is redundancy between sources $1$ and $2$. In particular, in $I(1;2|3)$, if variable $1$ as part of the target is associated with source $1$, then the contribution of $I(1;2|3)$ to $I(123;2)$ can be interpreted as redundant with the information $I(123;1) = I(1;1)$ in Eq.\,\ref{r33} that source $1$ has about itself. The second decomposition in Eq.\,\ref{r34}b further challenges the interpretation of redundancy and unique information based on the assignment of an identity to bits of information in the target given their association with the overlapping target variables. Given $I(3;2|1)$, source $2$ provides information about $3$. But the amount of information contained in $3$ is shared with $1$ and $2$, given the conditional dependencies of the XOR system. Moreover, in $I(3;2|1)$ the conditioning on variable $1$ as a target, if this variable is associated with source $1$, would suggest that $I(3;2|1)$ contributes information to $I(123;2)$ by combining the two sources. Accordingly, when using the target-sources correspondence to identify pieces of information, different partitioning orders of the mutual information ambiguously suggest that information can be obtained uniquely, redundantly, or even in a synergistic way. These problems arise because, in contrast to the case of $I(12;1,2)$ with independent sources, in the XOR system the two bits of $123$ cannot be identified as belonging to a certain variable, but can only be distinguished as the bit that any first variable provides alone, and the bit that a second variable provides combined with the first. This lack of correspondence between pieces of information and individual variables is incompatible with the identification of the pieces of information based on the association of target-sources overlapping variables.

The differences in the quantification of redundancy with each stochasticity axiom are related to the alternative interpretations of identity discussed for Eqs.\,\ref{r33} and \ref{r34}. A notion of redundancy compatible with the weak axiom considers the common information about the target that can be obtained by observing either source alone or conditioned on variables in the target. Indeed, the deterministic component of redundancy comprises the conditional dependence of the sources given the rest of the target, $I(1;2|X \backslash 12)$, when there is a target-sources overlap, and thus fits to Eq.\,\ref{r34}a. Conversely, with the strong axiom, when there is a target-source overlap, redundancy equals $I(1;2)$ independently of $X \backslash 12$, in agreement with the logic of Eq.\,\ref{r33}. We will further discuss the implications of the axioms about information identity in Section \ref{s53} after dealing with the trivariate case.

\section{Trivariate decompositions with deterministic target-sources dependencies}
\label{s5}

We now extend the analysis to the trivariate case. This is relevant because, in contrast to the bivariate case, it has been argued that with more than two sources the PID decompositions that jointly comply with the monotonicity and the identity axiom do not guarantee the nonnegativity of the PID terms \citep{Bertschinger12b}. In particular, \cite{Bertschinger12b} used the XOR example we reconsidered above as a counterexample to show that negative terms appear. Therefore, we would like to be able to extend the general formulation of Section \ref{s41} to the trivariate case, and thus apply it to further examine the XOR and AND examples by identifying each component of the trivariate decomposition of $I(123;123)$ and not only of the decomposition of $I(123;12)$.

\subsection{General formulation}
\label{s51}

While in the bivariate decomposition there is a single PID term that involves synergistic information, in the trivariate lattice of Figure \ref{fig1}B all nodes which are not reached descending from $1$, $2$, or $3$ imply synergistic information, and the nodes of the form $i.jk$ too. The weak and strong axioms impose constraints on these terms given Eqs.\,\ref{r0} and \ref{r0b}, respectively.

\subsubsection{PID decompositions with the weak axiom}
\label{s511}

We start with the weak stochasticity axiom. Consider that any of the three primary sources is part of the target and how synergy may appear. For example, consider the extra information obtained when observing $12$ together instead of $1$ and $2$ separately. This information is distributed across the nodes reached descending from $12$ that are not already reached descending from $1$ or from $2$. But the axiom states that if any of $1$ and $2$ is contained in the target, $12$ cannot have synergistic information about them. Furthermore, if $3$ is part of the target and $12$ provides some extra information about it not given by $1$ and $2$ alone, this information is redundant with the one that $3$ provides about itself, and hence is contained in the node $3.12$, which is still reachable descending from $3$. Accordingly, the weak stochasticity axiom implies that

\begin{equation}
\Delta_d(X; \alpha) =0\ \ \forall \alpha \notin \bigcup_{i=1,2,3} \downarrow i.
\label{r15}
\end{equation}

We can then proceed analogously to the bivariate case to characterize the remaining deterministic contributions to PID terms. We again apply the mutual information chain rule to separate stochastic and deterministic dependencies. Again we focus on the partitioning order that considers first the stochastic dependencies, since only this order leads to an additive separation of stochastic and deterministic components for each PID term. With this partitioning order

\begin{equation}
\begin{split}
I(X;123) &=  I(X \backslash 123;123)+ I(X \cap 123; 123| X \backslash 123)\\
&= I(X \backslash 123;123)+ H(X \cap 123| X \backslash 123).
\label{r15b}
\end{split}
\end{equation}
Following derivations analogous to the ones of Section \ref{s41} (see Appendix \ref{a1}), it can be seen that deterministic contributions are further restricted by

\begin{equation}
\Delta_d(X; \alpha) =0\ \ \forall \alpha \notin \bigcup_{i \in X \cap \{1, 2, 3\}} \downarrow i.
\label{r15c}
\end{equation}
If a certain primary source $i$ does not overlap with the target, the nodes that can only be reached descending from its corresponding node will not have a deterministic component. This can be understood intuitively. For example, suppose that the target includes $1$ and $2$ but not $3$. Then the entropy in Eq.\,\ref{r15b} is $H(12|X \backslash 123)$. The PID terms that can be reached descending from $3$ and not from $1$ or $2$ are $\Delta(X;3)$ and $\Delta(X;3.12)$ (see Figure \ref{fig1}B). The first quantifies information that can only be obtained from $3$, and not from $12$. The second is information that can be obtained from $3$ or from $12$, but not from $1$ or $2$ alone. But since all the information about $12$ can be obtained either from $1$ or $2$, these nodes do not contribute to the decomposition of $H(12|X \backslash 123)$.

\begin{table}[!htbp]
\centering
\begin{tabular}{| l c | c |}
\hline \hline
Term &  Measure\\
\hline \hline
$\Delta_d(X; i)$ & $H(i|jk, X \backslash ijk)$ \\
\hline
$\Delta_d(X; i.jk)$ & $I(X \backslash jk; jk \backslash j,k) - I(X \backslash ijk; jk \backslash j,k)$ \\
\hline
$\Delta_d(X; i.j)$ & $I(i;j|k, X \backslash ijk)- \left [ \Delta_d(X; i.jk) + \Delta_d(X; j.ik) \right ]$ \\
\hline
$\Delta_d(X; i.j.k)$ & $C(i;j;k| X \backslash ijk) + \Delta_d(X; i.jk)+\Delta_d(X; j.ik)+\Delta_d(X; k.ij)$\\
\hline
\end{tabular}
\caption{Deterministic components of the PID terms for the trivariate decomposition derived from the weak stochasticity axiom. All terms not included in the table have no deterministic component due to the axiom. These expressions correspond to the case in which the primary source $i$ overlaps with the target. If $i$ does not overlap, $\Delta_d(X; i)$ and $\Delta_d(X; i.jk)$ are zero, while the other terms depend on their characteristic symmetry for the other variables $j$ and $k$, and cancel if none of the variables with the corresponding symmetry overlaps with the target. See the main text and Appendix \ref{a1} for details.}
\label{tab2}
\end{table}

Using the condition of Eq.\,\ref{r15c}, we can use the same procedure as in Section \ref{s41} to derive the expressions of all the deterministic PID trivariate components. These terms are collected in Table \ref{tab2} and we leave the detailed derivations and discussion for Appendix \ref{a1}. Their expressions are indicated for the case in which variable $i$ is part of the target and are symmetric with respect to $j$ or $k$ when this symmetry is characteristic of a certain PID term, or cancel otherwise, consistently with Eq.\,\ref{r15c}.

The first two terms $\Delta_d(X;i)$ and $\Delta_d(X;i.jk)$ are nonnegative, the former because it is an entropy and the latter because according to the axiom adding a new source can only reduce synergy. But for the terms $\Delta_d(X; i.j)$ and $\Delta_d(X; i.j.k)$ it is not guaranteed that they are nonnegative. For $\Delta_d(X; i.j)$, we will see examples of negative values below. For $\Delta_d(X; i.j.k)$, the conditional co-information can be negative if there is synergy between the primary sources when conditioning on the non-overlapping target variables, and this can happen when there is no synergy about the target, leading to a negative value. Therefore, following the weak stochasticity axiom, the PID decomposition cannot ensure the nonnegativity of all terms when deterministic target-sources dependencies are in place. We will further discuss this limitation after examining the full trivariate decomposition for the XOR and AND examples.

\subsubsection{PID decompositions with the strong axiom}
\label{s512}

With the strong form of the axiom, not only deterministic but stochastic components of synergy are restricted. There cannot be any synergistic contribution that involves a source overlapping with the target. Eq.\,\ref{r1b} can be applied with $S = 123$. Furthermore, since the cancelation of synergistic terms has to hold not only for the terms $\Delta(X; \alpha)$ of the trivariate lattice but also of any bivariate lattice associated with it, given the mapping of PID terms between these lattices (Figure \ref{fig1}), this implies that in the trivariate lattice also the PID terms of the form $i.jk$ are constrained. In fact, there is only one case in which synergistic contributions can be nonzero if there is any target-sources overlap for the trivariate case, and this is when only one variable overlaps. Consider that only variable $1$ is part of the target. Since there cannot be any synergy involving $1$, all synergistic PID terms contained in $I(X;1|2)$, $I(X;1|3)$, or $I(X;1|23)$ have to cancel, and also $\Delta(X; 2.13)$ and $\Delta(X; 3.12)$. It can be checked that this includes all synergistic terms except $\Delta(X; 23)$ and $\Delta(X; 1.23)$. The former quantifies synergy about other target variables and the latter synergy redundant with the information of $1$ itself. With more than one primary source overlapping with the target all synergistic terms have to cancel for the trivariate case.

\begin{table}[!htbp]
\centering
\begin{tabular}{| l c | c |}
\hline \hline
Term &  Measure\\
\hline \hline
$\Delta_d(X; i)$ & $H(i|jk)$ \\
\hline
$\Delta_d(X; i.jk)$ & $I(i; jk \backslash j,k)$ \\
\hline
$\Delta_d(X; i.j)$ & $I(i;j|k)-\Delta_d(X; i.jk)$ \\
\hline
$\Delta_d(X; i.j.k)$ & $C(i;j;k)+ \Delta_d(X; i.jk)$\\
\hline
\end{tabular}
\caption{Deterministic components of the PID terms for the trivariate decomposition derived from the strong stochasticity axiom. All terms not included in the table have no deterministic component due to the axiom. Again, the expressions shown here correspond to the case in which the source $i$ overlaps with the target. For $\Delta_d(X; i.jk)$ we further consider that neither $j$ nor $k$ overlap with the target, and otherwise this term cancels. If $i$ does not overlap, $\Delta_d(X; i)$ is zero, while the other terms depend on their characteristic symmetry for the other variables $j$ and $k$ and cancel otherwise. See the main text and Appendix \ref{a1} for details.}
\label{tab3}
\end{table}

Like for the weak axiom, we now leave the derivations for Appendix \ref{a1}. The PID deterministic terms are collected in Table \ref{tab3}, again for simplicity showing their expressions for the case in which $i$ overlaps with the target. The form of the expressions respects the symmetries of each term. For example, if $j$ instead of $i$ overlaps with the target then $\Delta_d(X; i.j) = I(i;j|k)-\Delta_d(X; j.ik)$. Note however that, because $\Delta_d(X; j.ik)=0$ when $i$ overlaps, if both $i$ and $j$ overlap then $\Delta_d(X; i.j) = I(i;j|k)$. See Appendix \ref{a1} for further details.

In comparison to the deterministic components derived from the weak axiom there are two differences: First, the lack of conditioning on $X \backslash ijk$ is due to the reversed partitioning order selected. Like for the bivariate case, the deterministic PID components are independent of the non-overlapping target variables when adopting the strong stochasticity axiom. Second, assuming the strong axiom the terms $\Delta_d(X; i.jk)$ can only be nonzero if $j$ and $k$ are not contained in the target and when more than one source overlaps all terms of the form $\Delta_d(X; i.jk)$ cancel. In that case it is clear that $\Delta_d(X; i.j.k)$ can be negative, since the co-information can be negative. Therefore, also the PID decomposition derived from the strong axiom does not ensure nonnegativity. We will now show examples of negative terms for both PID decompositions.

\subsection{Illustrative systems}
\label{s52}

We now continue the analysis of the XOR and AND examples by decomposing $I(123;123)$. Since now $X \backslash 123 = \emptyset$ the decompositions are completely deterministic and are obtained calculating the PID components described in Table \ref{tab2} and Table \ref{tab3}. Accordingly, given that deterministic and joint PID terms are equal, we will use $\Delta(X; \beta)$ instead of $\Delta_d(X; \beta)$ to refer to them.

\subsubsection{XOR}
\label{s521}

\begin{figure}[!htbp]
  \begin{center}
    \scalebox{0.4}{\includegraphics*{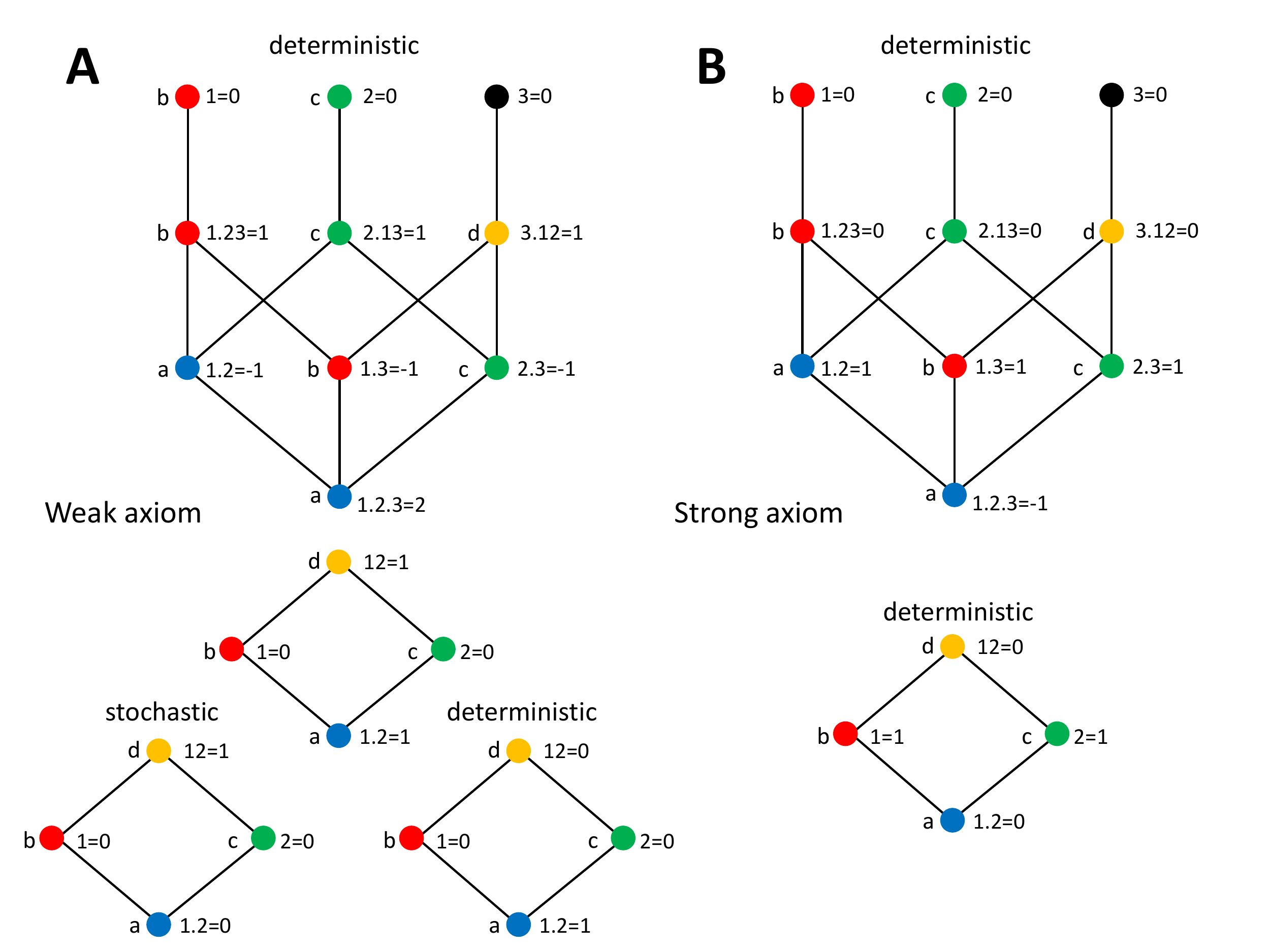}}
  \end{center}
  \caption{Trivariate decompositions of $I(123;123)$ for the XOR system. \textbf{A}) Decomposition derived from the weak stochasticity axiom. The trivariate redundancy lattice is displayed only for nodes lower than the single source nodes because all upper PID terms are zero. The bivariate decomposition of $I(123;12)$ is shown again now indicating the mapping of the PID terms with colors and labels as in Figure \ref{fig1}. \textbf{B}) Same as A) but for the decomposition derived from the strong axiom.}
  \label{fig3}
\end{figure}

We start with the XOR example and the decomposition derived from the weak stochasticity axiom (Figure \ref{fig3}A). We show the trivariate decomposition of $I(123;123)$ and also again the decomposition of $I(123; 12)$, now indicating the mapping of the nodes with the trivariate decomposition. For the trivariate lattice we only show the nodes lower than the ones of the primary sources because for all others the corresponding terms are zero (Eq.\,\ref{r15}). The PID terms are calculated considering Table \ref{tab2} and the information-theoretic quantities displayed in Figure \ref{fig2a}A.

The trivariate terms $\Delta(X; i)$ are all zero, because any two variables determine the third. This is also reflected in the terms $\Delta(X; i.jk)$ having $1$ bit. The terms $\Delta(X; i.j)$ are all equal to $-1$ bit. These terms should quantify the redundant information between two variables which is unique with respect to the third, but their interpretation is impaired by the negative values. Furthermore, $\Delta(X; i.j.k)=2$, so that redundancy monotonicity does not hold. However, it can be checked that the values obtained are consistent from the point of view of the constraints linking PID terms and mutual informations. Similarly, the calculated PID components are consistent between the bivariate and trivariate decompositions. In particular, the sum of the nodes with the same color or label in the trivariate lattice equals the corresponding node in the bivariate lattice. This equality holds for the joint bivariate lattice, and not for the deterministic lattice alone, even if in the trivariate case the lattice is uniquely deterministic. This reflects a transfer of stochastic synergy in the bivariate case to deterministic redundancy in the trivariate case (see yellow nodes labeled with $d$).

We now consider the decomposition derived from the strong axiom (Figure \ref{fig3}B). In this case also $\Delta(X; i)$ are all zero because any two variables determine the third, but now also $\Delta(X; i.jk)$ are zero. This is because the axiom assumes that there is no synergy involving any of the primary sources overlapping with the target. $\Delta(X; 3.12)=0$ is consistent with the lack synergy for the decomposition of $I(123;12)$, as indicated by the mapping of the yellow nodes labeled with $d$. Also the mapping of all other PID terms is consistent. In particular, the $1$ bit corresponding to the unique informations of the bivariate decomposition are contained in the terms $\Delta(X; i.j)= I(i;j|k)$ of the trivariate one. In comparison to the decomposition from the weak axiom, these terms are not negative, but instead a negative value is obtained for $\Delta(X; i.j.k)$. Therefore nonnegativity is neither fulfilled for this decomposition.

In fact, this decomposition was used by \cite{Bertschinger12b} and \cite{Rauh14} as a counterexample to show that with more than two sources there is no decomposition that can simultaneously comply with the redundancy monotonicity axiom and the identity axiom and also lead to global nonnegativity of the PID terms. In Section \ref{s411} we pointed out that the bivariate decomposition of $I(123;12)$ derived from the strong stochasticity axiom coincides with the one obtained from the arguments of \cite{Bertschinger12b} based on the identity axiom. However, the trivariate decomposition we obtain is not the same as they did. The divergence occurs because \cite{Bertschinger12b}, after arguing that $I(123; 1.2)=0$ based on the identity axiom as we discussed in Section \ref{s441}, further argued that this implies $I(123; 1.2.3)=0$, based on redundancy monotonicity, and hence also $\Delta(X; i.j)=0$. Once having all terms $\Delta(X; i.j)=0$ and $\Delta(X; i.j.k)=0$, this led them to find a negative value for $\Delta(X; 12.13.23)$. However, as we can see in the trivariate decomposition of Figure \ref{fig3}B, to respect the monotonicity axiom having $I(123; 1.2)=0$ does not necessarily imply that $\Delta(X; i.j)=0$ and $\Delta(X; i.j.k)=0$. Indeed, monotonicity is respected by already having a negative value $\Delta(X; i.j.k)=-1$, as we get from assuming the strong stochasticity axiom. Accordingly, once recognising this possibility, the results obtained from the strong axiom are compatible with the arguments of \cite{Bertschinger12b}, showing that this decomposition is a counterexample for global nonnegativity, but indicating that a negative value appears already in $\Delta(X; i.j.k)$.

\subsubsection{AND}
\label{s522}

\begin{figure}[!htbp]
  \begin{center}
    \scalebox{0.4}{\includegraphics*{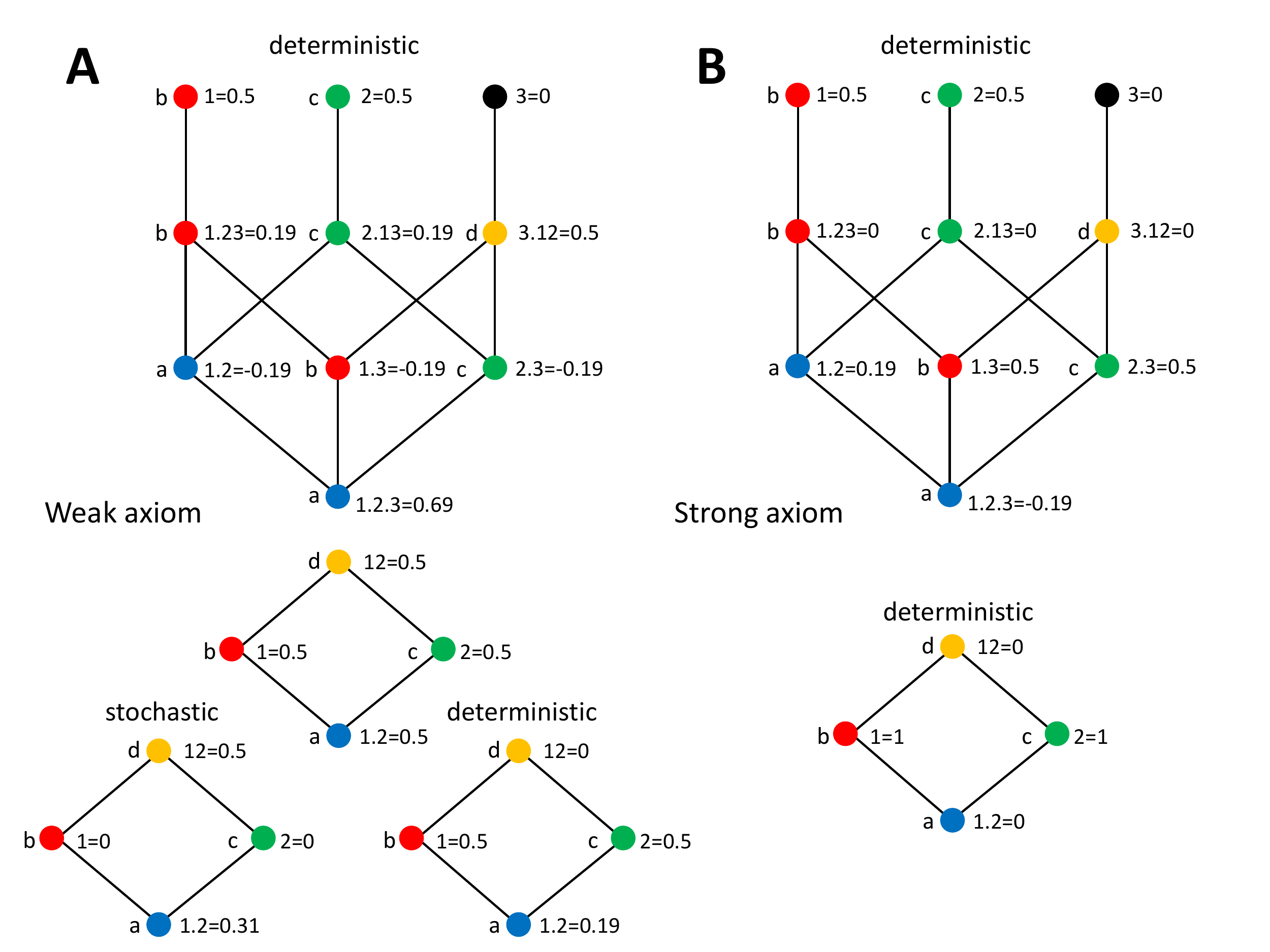}}
  \end{center}
  \caption{Trivariate decompositions of $I(123;123)$ for the AND system. The structure of the figure is the same as in Figure \ref{fig3}.}
  \label{fig4}
\end{figure}

We present the AND decomposition as a further example. All PID terms are derived using the information-theoretic quantities of Figure \ref{fig2b}A in combination with Tables \ref{tab2} and \ref{tab3}. Like for the XOR case, the mapping of trivariate to bivariate decompositions is consistent. Again, both trivariate decompositions contain some negative term. With the strong axiom, while the bivariate decompositions for the XOR and AND example are equal, the trivariate PID terms differ substantially, reflecting the different symmetries of each operation. In particular, for the XOR case there is a symmetry between all three variables while in the AND case only between the inputs.

\subsection{PID terms nonnegativity and Information identity}
\label{s53}

The analysis of the trivariate PID decompositions derived from the weak and strong versions of the stochasticity axiom shows explicitly how nonnegative PID terms can arise in the presence of deterministic target-sources dependencies. The form of the deterministic components indicated in Tables \ref{tab2} and \ref{tab3} provides a general understanding of how negative PID terms can occur, beyond the concrete counterexample examined in \cite{Bertschinger12b}. In particular, the axioms enforce that certain pieces of information are attributed to redundancy or unique information terms because their identity is associated to the sources, and hence deterministic components of the decomposition are bounded to the low part of the redundancy lattice, which leads to negative terms in order to conform to the lattice structure and to the relations between PID terms and mutual informations. Furthermore, as argued by \cite{Rauh17b} based on continuity arguments for the mutual information, the same problem of obtaining negative PID terms is expected to occur not only when deterministic target-sources dependencies exist, but also in the limit of strong dependencies tending to be deterministic.

Avoiding negative PID terms would require changing the assumptions about how deterministic target-sources dependencies constrain the terms. The common assumption of the weak and strong axioms that information about an overlapping variable can only be redundant or unique may be too restrictive and implies assuming that we can assign an identity to pieces of information in the target as exclusively related to the overlapping variable. Conversely, only in few cases the identity of a bit can be assigned to a single variable, as it is the case for $I(12;12)$ with $1$ and $2$ being independent, which motivated the identity axiom and in particular the Independent Identity property \citep{Ince16} that requires $I(12;1.2)=0$ for this case.

In general, the overall composition of the target affects the identity of each piece of information. For example, even if $1$ and $2$ are independent and for target $12$ we can identify each piece as associated with a different variable, if we incorporate a third variable $3$ determined by $1$ and $2$, now identity will generally change, and depend on the specific operation that generates $3$ from $1$ and $2$. This is the case, for example, of the XOR system when $123$ is taken as the target. The two bits of $123$ cannot be identified as belonging to a certain variable, but only as the bit that any first variable provides alone, and the bit that a second variable provides combined with the first. Oppositely, on one hand, the strong axiom assumes that each source alone can uniquely provide a bit, corresponding to its own identity, as reflected in the decomposition of $I(123;12)$ (see Figure \ref{fig3}B). On the other hand, with the weak axiom, the second bit is classified as synergy, consistently with the idea that retrieving it requires the combination of two variables (Figure \ref{fig3}A). However, because the weak axiom still assumes that any information about an overlapping variable has to be redundant or unique, it imposes that the synergy is contained in the terms $\Delta(X; i.jk)$ in the trivariate decomposition and not in terms corresponding to nodes upper than the ones of single variables. This means that the axiom is still not compatible with the identification of the two bits as the one that can be obtained from a single variable and the one that can only be obtained from the combination of two variables.

\begin{figure}[!htbp]
  \begin{center}
    \scalebox{0.4}{\includegraphics*{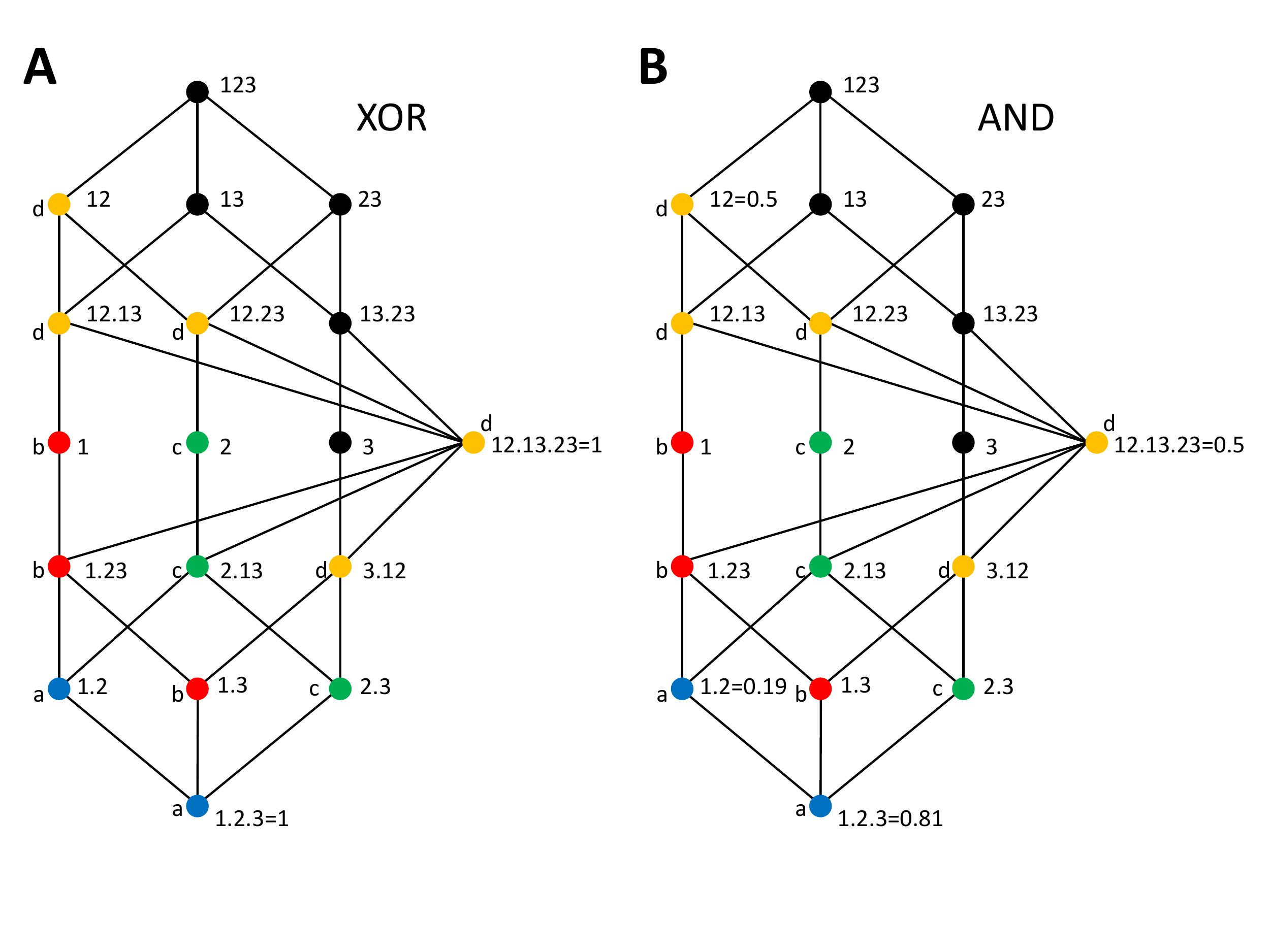}}
  \end{center}
  \caption{Nonnegative decompositions of $I(123;123)$ based on the identification of pieces of information in the target without imposing constraints to synergy due to deterministic target-sources dependencies. \textbf{A}) Decomposition for the XOR system. \textbf{B}) Decomposition for the AND system. We keep the colors and labels of the nodes to facilitate the comparison with Figures \ref{fig3} and \ref{fig4}. Terms for which the value is not indicated are zero.}
  \label{fig6}
\end{figure}

In Figure \ref{fig6}A we show a trivariate decomposition that is consistent with this identification of the bits for the XOR example. The first bit is the one that can be obtained from any variable alone, and is thus redundant to all three variables. None of the variables can provide more information alone, so all the remaining PID terms associated with nodes lower than $1$, $2$, and $3$ are zero. Furthermore, since the second bit can be obtained by the combination of any two variables, its information is redundant to any pair, and is thus contained in the node $12.13.23$. Since in total there are two bits the rest of PID terms are also zero. This decomposition is nonnegative, but does not conform to any version of the stochasticity axiom (neither to the identity axiom) because $\Delta(X; 12.13.23)>0$. In fact, it corresponds to the one obtained using as redundancy measure $\mathrm{min} \{ I(X; A_i)\}$ over the sources $A_i$ of each collection \citep{Bertschinger12b}, which is closely related to the measure of \cite{Williams10}.

With this same measure we also get a nonnegative decomposition for the AND example (Figure \ref{fig6}B). Like for the XOR case, when we move from target $12$ to target $123$ there is no qualitative argument to associate the two bits of the target to particular source variables. Identity of the pieces of information is assigned only with a quantitative criterion: Since $H(3)\approx 0.81$ this is the maximum information that can be redundant to all three sources about the target $123$. Both $1$ and $2$ can still provide alone $\approx 0.19$, reaching one bit of information. Any combination of two primary sources provides another half bit. In total this provides already the information corresponding to the entropies $H(13)$ and $H(23)$, which is one and a half bits. The remaining information is unique of the synergystic term combining $12$.

Overall, this analysis highlights that distinguishing redundant, unique, and synergistic information requires a criterion to assign identity to each piece of information in the target. For measures fulfilling the identity axiom \citep{Bertschinger12} and more generally complying with the stochasticity axioms, the implicit criterion assumes that the identity of the sources is preserved within the target. This criterion respects the Independent Identity property but leads to negative terms. Oppositely, the criterion is quantitative for the redundancy measure used for Figure \ref{fig6} or for the measure of \cite{Williams10} and while it leads to a nonnegative decomposition it does not respect the intuition about qualitative redundancy of the Independent Identity property. It remains an open question if there is a general criterion of identity that accommodates the intuition associated with the Independent Identity property and is also compatible with the relations intrinsic to the redundancy lattice of \cite{Williams10}.

\section{Discussion}

\subsection{Implications for the theoretical definition of redundant, synergistic and unique information}

The proposal of \cite{Williams10} of decomposing mutual infor¬mation into nonnegative redundant, unique, and synergistic components has been a fruitful and influential conceptual framework. However, a concrete implementation consistent with a set of axioms formalizing the notions for such types of information has proven to be elusive. The main difﬁculty stems from determining if redundant sources contain the same qualitative information, which requires assigning an identity to pieces of informa¬tion in the target. \cite{Harder12} pointed out that the redundancy defined by \cite{Williams10} only captures quantitatively the common amounts of information shared by the sources. They introduced the identity axiom to ensure that two independent variables cannot have redundant information about a copy of themselves. The lack of redundancy for this particular case has been enunciated as the Independent Identity property by \cite{Ince16}. However, \cite{Bertschinger12b} provided a counterexample showing that nonnegativity of the PID terms is not ensured when the identity axiom is assumed. This counterexample also involved a target constituted as a copy of the primary sources, in particular as the inputs and output variables of the XOR logical operation. Since these cases in which a deterministic relation exists between the target and the sources have played an important role both in motivating the identity axiom and raising caveats about the internal consistency of the axioms, in this work we have examined more systematically the effect of deterministic target-sources dependencies in the PID decomposition.

We introduced (Section \ref{s3}) two variants of a stochasticity axiom that imposes some constraints to the existence of synergy in the presence of deterministic target-sources dependencies. In presence of a target-sources overlap, the weak form of the axiom states that there cannot be synergistic information about the overlapping target variables. The strong axiom further constrains synergy assuming that the overlapping sources cannot provide other information than about themselves, and thus can neither contribute synergistic information about the non-overlapping part of the target.

We derived (Section \ref{s41}) general formulas for the PID terms in the bivariate case, following each version of the stochasticity axiom. We showed that the PID terms can be separated into a stochastic and a deterministic component, which account for the information about the non-overlapping and overlapping target variables, respectively. We indicated that the stochasticity axioms subsume the identity axiom and provide two alternative generalizations to characterize redundancy for any multivariate system with any degree of target-sources overlap (Section \ref{s42}). We checked how several previously proposed measures conform to these generalizations (Section \ref{s43}). We then examined (Section \ref{s44}) two concrete examples based on the XOR and AND logical operations, with variables $1$ and $2$ as inputs and variable $3$ as output, calculating the PID decomposition of the mutual information $I(123;12)$.

Based on these examples we argued that the two axioms imply different interpretations of redundancy as common information about the target that can be obtained by observing either source alone. With the weak axiom each source can be combined with some target variables to provide information about other target variables, even in the presence of a target-sources overlap. Conversely, the strong axiom assumes that any overlapping variable only provides information about itself, and thus redundant information cannot appear because a source is combined with target variables. This leads to redundancy being always equal to the mutual information between the primary sources when there is some target-sources overlap, independently of the non-overlapping target variables. However, while enforcing different constraints, both stochasticity axioms assume that the identity of the overlapping sources is preserved within the target, that is, that some pieces of information have an identity that can be associated only with the corresponding sources, independently of the other variables composing the target.

In Section \ref{s51} we extended the general derivations to the trivariate case. This allowed us to understand what originates negative PID terms -as found in the counterexample of \cite{Bertschinger12b}- from a general perspective. We identified that, following the stochasticity axioms, several PID terms have a deterministic component that is not nonnegatively defined. We resumed the XOR and AND examples in Section \ref{s52}, now analyzing the decomposition of $I(123;123)$, as was done in the counterexample of \cite{Bertschinger12b}. \cite{Bertschinger12b} did not fully characterize this decomposition, but instead used the identity axiom to indicate that at least a negative PID term appears. We confirmed their results indicating that their arguments are consistent with the strong axiom, although we pointed out that a negative value is already present for the redundancy between the three variables, instead of in an upper PID term involving synergy.

Our analysis applying the general derivations from the stochasticity axioms allowed us to expose the relation between the assumptions on information identity and the lack of nonnegativity. In particular, imposing that certain pieces of information can only be attributed to redundancy or unique information terms based on the premise that their identity is associated to the sources enforces that deterministic components of the mutual information are bounded to the low part of the redundancy lattice, and this leads to negative terms in order to conform to the lattice structure and to the relations between PID terms and mutual informations.

The identity axiom was motivated by the especial case of $I(12;1, 2)$ with $1$ and $2$ being independent, for which, as particularly enunciated in the Independent Identity property \citep{Ince16}, redundancy should intuitively cancel because the pieces of information in the target can be separately assigned to each source. However, in general, the overall composition of the target affects the identity of each piece of information. For example, even if $1$ and $2$ are independent, incorporating to the target a third variable $3$ determined by $1$ and $2$ alters the identification of the target information. This is the case for the XOR example with the target formed jointly by the inputs and output variables, since the two bits of $123$ cannot be identified as belonging to any of the three variables, but only as the bit that any first variable provides alone, and the bit that a second variable provides combined with the first. In Section \ref{s53} we examined an alternative decomposition consistent with this alternative identification of the two bits of the XOR system. We showed that, for both the XOR and AND example, nonnegative decompositions are attained by admitting nonzero synergistic contributions. However, the identity criterion used in this alternative decomposition is purely quantitative, as the one of \cite{Williams10}, and thus does not respect the desired Independent Identity property.

Although the notion of redundancy as information shared about the same pieces of information is intuitive in plain language, its precise implementation within the information-theoretic framework is not straightforward. The measure of mutual information has applications in many fields, such as communication theory and statistics \citep{Cover06}. Accordingly, a certain decomposition in terms of redundant, unique, and synergistic contributions may be compatible only with one of its interpretations. Indeed, if information is understood in the context of a communication channel \citep{Shannon48}, nonnegativity is required from its operational interpretation as the number of messages that can be transmitted without errors. Furthermore, semantic content cannot be attributed, and thus information identity should rely only on the statistical properties of the distribution of the target variables. For example, in the case of the target composed by two independent variables, identity is assigned based on independence. Alternatively, if mutual information is used as a descriptor of statistical dependencies \citep{Kullback1959}, nonnegativity is not required since locally negative information, or misinformation \citep{Wibral14b}, simply reflects a certain change in the probability distribution of one variable due to conditioning on another variable. With this interpretation of information based on local dependencies, a criterion of information identity can introduce semantic content in association with the specific value of the variables and common information of two sources can be associated with dependencies that induce coherent modifications of the probability distribution of the target variables \citep{Ince16}. These local measures of information may be interpreted operationally in terms of changes in beliefs, or in relation to a notion of information more associated with ideal observer analysis than with communication theory \citep{Wibral14b, Thomson05}. In this work, we have not considered local versions of mutual information, and we adopted the premise that nonnegativity is a desirable property for the PID terms.

\subsection{Implications for studying neural codes}

Determining the proper criterion of information identity to evaluate when information carried by different sources is qualitatively common is essential to interpret the results of the PID decomposition in practical applications, such as in the analysis of the distribution of redundant, unique, and synergistic information in neural population responses. For example, when examining how information about a multidimensional sensory stimulus is represented across neurons, the decomposition should identify information about different features of the stimulus, and not only common amounts of information. The PID terms should reflect the functional properties of the neural population so that we can properly characterize the neural code. On the other hand, nonnegativity of the PID terms facilitates their interpretation not only as a description of statistical dependencies, but as a breakdown of the information content of neural responses, for example to assess the intersection information between sensory and behavioural choice representations \citep{Panzeri17, Pica17, Pica17b}.

The underlying criterion of information identity for the PID decomposition is also important when examining information flows among brain areas because, only if redundant and unique information terms correctly separate qualitatively the information, we can interpret the spatial and temporal dynamics of how unique new information is transmitted across areas. It is common to apply dynamic measures of predictability such as Granger causality \citep{Granger69} to characterize information flows between brain areas \citep{Wibral14}. The effect of synergistic and redundant information components in the characterization of information flows with Granger causality has been studied \citep{Stramaglia14, Stramaglia16}, and \cite{Williams11} applied their PID framework to decompose the information-theoretic measure of Granger causality, namely Transfer entropy \citep{Marko73, Schreiber00b}, into terms separately accounting for state-independent and state-dependent components of information transfer. Furthermore, they also indicated which terms of the PID decompositions can be associated with information uniquely transmitted at a certain time or information transfer about a specific variable, such as a certain sensory stimulus \citep{Beer15}. These applications of the PID framework identify meaningful PID terms based on the redundancy lattice, and thus can be applied for any actual definition of the measures, but our considerations highlight the necessity to properly determine information identity in order to fully exploit their explanatory power.

Furthermore, our discussion of how the interpretation of information identity depends on the dependencies between the variables composing the target indicates that the analysis of how redundant, unique, and synergistic information components are distributed across neural population responses can be particularly useful in combination with interventional approaches \citep{Panzeri17, Chicharro14}. In particular, the manipulation of neural activity with optogenetics techniques \citep{Oconnor13, Otchy15} can disentangle causal effects from other sources of dependencies such as common factors. Although this work illustrates the principled limitations of current PID measures, their combination with these powerful experimental techniques can help to better probe the functional meaning of the PID terms.

\subsection{Concluding remarks}

Overall, we have studied the effect of deterministic target-sources dependencies in the PID decomposition by enunciating two variants of a new stochasticity axiom, comparing them to the identity axiom \citep{Harder12}, and discussing their implications regarding information identity. Our analysis suggests that, if the redundancy lattice of \cite{Williams10} is to remain as the backbone of a nonnegative decomposition of the mutual information, a new criterion of information identity should be established that, while conforming to the Independent Identity property, it is less restrictive in the presence of deterministic target-source dependencies than the ones underlying these axioms.

\paragraph{\textbf{Acknowledgments}}

This work was supported by the Fondation Bertarelli.

\paragraph{\textbf{Authors Contribution}}

All authors contributed to the design of the research. The research was carried out by Daniel Chicharro. The manuscript was written by Daniel Chicharro with the contribution of Stefano Panzeri and Giuseppe Pica. All authors have read and approved the final manuscript.

\paragraph{\textbf{Conflicts of Interest}}

The authors declare no conflicts of interest.

\vspace*{7mm}

\appendix{}

\section{Alternative partitioning orders for the bivariate decomposition with target-sources overlap}
\label{a0}

We here derive in more detail the alternative expressions for the unique and redundant information terms collected in Table \ref{tab1}, which are obtained applying the other mutual information partitioning order of Eq.\,\ref{r3}b. Using the relation decomposing conditional mutual information into unique information and synergy we get
\begin{equation}
\begin{split}
&I(X; 1 \backslash 2) = I(X; 1|2) - I(X; 12 \backslash 1,2)\\
&= I(X \cap 12 ;1|2) + I(X \backslash 12; 1|2,X \cap 12)- I(X \backslash 12; 12 \backslash 1,2).
\label{r8}
\end{split}
\end{equation}
This leads to express the unique information of $1$ as
\begin{equation}
I(X; 1 \backslash 2) =  \begin{cases} I(X \backslash 12; 1 \backslash 2)\ \  \mathrm{if}\ X \cap 1 = \emptyset\\ H(1|2)- I(X \backslash 12; 12 \backslash 1,2)\ \  \mathrm{if}\ X \cap 1 \neq \emptyset \end{cases}.
\label{r9}
\end{equation}
In this case the unique information is separated into nonadditive terms and involves the synergy about $X \backslash 12 $. This cross-over may seem at odds with the expression obtained with the other partitioning order (Eq.\,\ref{r5}), but on the contrary it reflects the internal consistency of the relations between the information-theoretic quantities: Eqs.\,\ref{r5} and \ref{r9} coincide if $1$ is not part of the target. For $X \cap 1 \neq \emptyset$, their equality
\begin{equation}
H(1|2)- I(X \backslash 12; 12 \backslash 1,2)= I(X \backslash 12 ;1 \backslash 2)+ H(X \cap 1| 2, X \backslash 12)
\label{r10}
\end{equation}
is consistent with the definition $I(X \backslash 12; 1|2) = H(1|2) - H(1| 2, X \backslash 12)$, taking into account that conditional information is the sum of the unique and synergistic components.

Proceeding as with the other partitioning order, once we have the expression of the unique information we can use the relation with the mutual information to determine redundancy:
\begin{equation}
I(X; 1.2) = \begin{cases} I(X \backslash 12; 1.2) \ \  \mathrm{if}\ X \cap 12 = \emptyset  \\ I(1;2)+ I(X \backslash 12; 12 \backslash 1,2) \ \  \mathrm{if}\ X \cap 12 \neq \emptyset \end{cases}.
\label{r12}
\end{equation}
Also here internal consistency with Eq.\,\ref{r7} holds. In particular, the equality
\begin{equation}
I(1;2)+ I(X \backslash 12; 12 \backslash 1,2) = I(X \backslash 12; 1.2) + I(1;2|X \backslash 12)
\label{r13}
\end{equation}
reflects that
\begin{equation}
\begin{split}
C(X \backslash 12 ; 1; 2) &=  I(1;2)- I(1;2|X \backslash 12)\\ &=  I(X \backslash 12; 1.2)-I(X \backslash 12; 12 \backslash 1,2)
\label{r14}
\end{split}
\end{equation}
because the co-information is invariant to permutations (Eq.\,\ref{e5}) and also corresponds to the difference of the redundancy and synergistic PID components.

Also following the strong axiom the alternative partitioning order, in this case the one considering first stochastic dependencies with the non-overlapping target variables, can be derived. With overlap, Eq.\,\ref{r2b} implies that $I(X; 1 \backslash 2)= I(X; 1 | 2)$. For the unique information we get

\begin{equation}
I(X; 1 \backslash 2) = I(X \backslash 12 ;1 \backslash 2) + I(X \backslash 12 ;12 \backslash 1, 2) + H(X \cap 1| 2, X \backslash 12),
\label{r29}
\end{equation}
and for the redundancy

\begin{equation}
I(X; 1.2) = I(X \backslash 12; 1.2) + I(1;2|X \backslash 12) - I(X \backslash 12; 12 \backslash 1, 2).
\label{r30}
\end{equation}
Like with the weak axiom, internal consistency holds for the expressions obtained with the two partitioning orders.

\section{Derivations of the trivariate decomposition with target-sources overlap}
\label{a1}

We here derive in more detail the trivariate deterministic PID components. We start with the derivations following the weak stochasticity axiom. If we consider the unique information of one primary source with respect to the other two, for example $I(X; 3 \backslash 12)$, we have that

\begin{equation}
\begin{split}
I(X; 3 \backslash 12) &= \Delta(X;3) \\ &= I(X; 3|12)- \left[ \Delta(X;123)+\Delta(X;13)+\Delta(X;23)+\Delta(X;13.23) \right].
\label{r16}
\end{split}
\end{equation}
The weak axiom imposes for the trivariate case that synergy deterministic components upper than the single source nodes have to be zero (Eq.\,\ref{r15}). Accordingly, any deterministic component of $I(X; 3|12)$ has to be contained in $\Delta(X;3)$. Decomposing this conditional mutual information with the partitioning order that considers first the dependencies with the non-overlapping target variables

\begin{equation}
I(X; 3 \backslash 12) = I(X \backslash 123; 3 \backslash 12)+ H(X \cap 3|12, X \backslash 123),
\label{r17}
\end{equation}
and thus in general

\begin{equation}
\Delta_d(X;i) = H(X \cap i|jk, X \backslash ijk).
\label{r18}
\end{equation}

We now consider the conditional information of two primary sources given the third, for example

\begin{equation}
\begin{split}
I(X;23|1) &= I(X \backslash 123; 23|1) + H(X \cap 23| 1 , X \backslash 123).
\label{r19}
\end{split}
\end{equation}
The deterministic part $H(X \cap 23| 1, X \backslash 123)$ again can only be contained in the PID terms contributing to $I(X; 23|1)$ that are lower than the single source nodes. This means that it has to be contained in the terms
\begin{equation}
\begin{split}
\Delta_d(X;2) +\Delta_d(X;3)+\Delta_d(X;2.3)+\Delta_d(X;3.12)+\Delta_d(X;2.13).
\label{r19b}
\end{split}
\end{equation}
Furthermore, this conditional entropy can be decomposed considering explicitly the part of the uncertainty associated with conditional entropies of the form of Eq.\,\ref{r18}:

\begin{equation}
\begin{split}
&H(X \cap 23| 1, X \backslash 123) = H(X \cap 3| 1, X \backslash 123)+ H(X \cap 2| 1, X \cap 3, X \backslash 123) \\ &= H(X \cap 3| 12, X \backslash 123) + I(2; X \cap 3)|1, X \backslash 123)+ H(X \cap 2| 1, X \cap 3, X \backslash 123).
\label{r20}
\end{split}
\end{equation}
Accordingly, using the definition of the terms $\Delta_d(X;i)$ in Eq.\,\ref{r18} and combining Eqs.\,\ref{r19} and \ref{r19b} we get the following equalities. First,

\begin{equation}
\Delta_d(X;i)+ \Delta_d(X;i.j) + \Delta_d(X;i.jk)+ \Delta_d(X;j.ik) = H(i|k, X \backslash ijk)\ \mathrm{if}\ X \cap i \neq \emptyset ,
\label{r21}
\end{equation}
and second

\begin{equation}
\Delta_d(X;i.j) + \Delta_d(X;i.jk)+ \Delta_d(X;j.ik) = I(i; j|k, X \backslash ijk)\ \ \mathrm{if}\ X \cap i \neq \emptyset.
\label{r22}
\end{equation}
Like in the expressions of the deterministic PID components in the Tables of Sections \ref{s41} and \ref{s51}, we here for simplicity indicate the equalities that hold when the primary source $i$ overlaps with the target. The symmetries of each $\Delta_d(X; \beta)$ term indicate when it can be nonzero. For example, $\Delta_d(X; i.j)$ is constrained by an equality of the form of Eq.\,\ref{r22} both if $i$ or $j$ overlap with the target.

Finally, we consider also how an unconditional mutual information is decomposed in PID terms. For example, again using the partitioning order that considers first stochastic target-source dependencies we have

\begin{equation}
\begin{split}
I(X;3) &= I(X \backslash 123;3)+ I(X \cap 123; 3|X \backslash 123) \\
&= I(X \backslash 123;3)+ H(X \cap 3|X \backslash 123)\ \ \mathrm{if}\ X \cap 3 \neq \emptyset.
\label{r23}
\end{split}
\end{equation}
When $3$ is part of the target the deterministic part of this information has to be contained in the nodes reached descending from $3$, and thus in general

\begin{equation}
\sum_{\beta \in \downarrow i} \Delta_d(X;\beta)= H(i|X \backslash ijk)\ \ \mathrm{if}\ X \cap i \neq \emptyset .
\label{r24}
\end{equation}
Combining Eq.\,\ref{r24} with Eq.\,\ref{r21} we get that

\begin{equation}
\begin{split}
\Delta_d(X;i.j) + \Delta_d(X;i.j.k)- \Delta_d(X;k.ij) &= I(i;j|X \backslash 123) \ \ \mathrm{if}\ X \cap i \neq \emptyset \\ &= H(i|X \backslash 123)-H(i|j, X \backslash 123).
\end{split}
\label{r25}
\end{equation}
Altogether, from Eqs.\,\ref{r18}, \ref{r21}, \ref{r22}, \ref{r24}, and \ref{r25} we can proceed to obtain expressions of the PID terms as a function of mutual informations and entropies. Doing so, the rest of PID terms remain as a function also of the terms $\Delta_d(X; i.jk)$. These terms can be understood by comparing the trivariate decomposition and a bivariate decomposition with only sources $j$ and $k$. For the latter, if $i$ is part of the target, $I(i; jk \backslash j,k)$ quantifies a stochastic synergistic contribution, because $i$ is not a source. Conversely, in the trivariate decomposition $i$ is a source and this information is now redundant with the information provided by variable $i$ itself. This means that we can identify $\Delta_d(X;i.jk)$ by comparing synergy between these two decompositions. For example, for the bivariate decomposition of $I(X; 12)$, $3$ is not a source and according to the weak axiom synergy can provide information about the non-overlapping part of the target, which can comprise $3$. Moving to the trivariate case by adding $3$ as a source this synergy stochastic component becomes redundant to information source $3$ has about itself, and thus

\begin{equation}
\begin{split}
I(X; 12 \backslash 1,2) &= I(X \backslash 12; 12 \backslash 1,2) \\
&= I(X \backslash 123; 12 \backslash 1,2) + \left [ I(X \backslash 12; 12 \backslash 1,2) - I(X \backslash 123; 12 \backslash 1,2) \right ].
\end{split}
\label{r26}
\end{equation}
In general, this means that these type of PID terms can be quantified as

\begin{equation}
\begin{split}
\Delta_d(X; i.jk) =  I(X \backslash jk; jk \backslash j,k) - I(X \backslash ijk; jk \backslash j,k).
\end{split}
\label{r27}
\end{equation}
These terms are nonnegatively defined, because according to the axiom adding a new source can only reduce synergy. After calculating these terms we can obtain all the expressions collected in Table \ref{tab2}.

For the strong stochasticity axiom, instead of repeating all the derivations we proceed by arguing about what has to change with respect to the decomposition obtained for the weak axiom. Changes originate from the difference in the constraints that both versions of the axiom impose on the existence of synergistic components and from the alternative mutual information partitioning order that leads to an additive separation of stochastic and deterministic PID components depending on the axiom. With the strong axiom this additive separation is reached using the partitioning order that first considers deterministic target-sources dependencies. This means that the conditioning of entropies and mutual informations on $X \backslash ijk$ will in this case not be present. Moreover, since the strong axiom restricts also synergy with the non-overlapping target variables, even a stochastic component of $I(X; 12 \backslash 1,2)$ can only be nonzero if $3$, but neither $1$ or $2$, overlap with the target. Since once further adding $3$ to the sources any synergistic component should be zero, the expression of the terms $\Delta_d(X; i.jk)$ in Eq.\,\ref{r27} is reduced to $I(i; jk \backslash j, k)$ when only $i$ overlaps with $X$, and to zero otherwise. Implementing these two modifications, the expressions of Table \ref{tab3} are obtained from the ones of Table \ref{tab2}.


\end{document}